\documentclass[fleqn,10pt]{article}
\usepackage[utf8]{inputenc}
\usepackage{amssymb,epsf,epsfig,graphicx,float,mathabx}
\usepackage[dvips]{color}
\usepackage{authblk}
\usepackage[compress]{natbib}
\setcitestyle{super,open={},close={},comma}
\usepackage[french,english]{babel}
\usepackage{bbm}

\usepackage{multicol}
\makeatletter
\newdimen\rh@wd
\newdimen\rh@hta
\newdimen\rh@htb
\newbox\rh@box
\def\rh@measure#1{\setbox\rh@box=\hbox{$#1$}\rh@wd=\wd\rh@box \rh@hta=\ht\rh@box}

\def\widecheck#1{\rh@measure{#1}%
  \setbox\rh@box=\hbox{$\widehat{\vrule height \rh@hta width\z@ \kern\rh@wd}$}%
  \rh@htb=\ht\rh@box \advance\rh@htb\rh@hta \advance\rh@htb\p@
  \ooalign{$\vrule height \ht\rh@box width\z@ #1$\cr
           \raise\rh@htb\hbox{\scalebox{1}[-1]{\box\rh@box}}\cr}}
\makeatother

\makeatletter
\def\smallunderbrace#1{\mathop{\vtop{\m@th\ialign{##\crcr
   $\hfil\displaystyle{#1}\hfil$\crcr
   \noalign{\kern3\p@\nointerlineskip}%
   \tiny\upbracefill\crcr\noalign{\kern3\p@}}}}\limits}
\makeatother
\newcommand{\ii}{\mathbbm{i}}
\newcommand{\e}{\mathop{\,\rm e}\nolimits}

\newcommand{\gb}{\mathop{g_{\rm b}}}
\newcommand{\gf}{\mathop{g_{\rm f}}}
\newcommand{\geul}{\mathop{\gamma_{\atop\hspace{-0.05cm}\rm E\!\!}}}
\newcommand{\dig}{\mathop{\psi_{\atop\!\!\!\!\rm dig\!\!}}}
\newcommand{\R}{\mathbb{R}}
\newcommand{\Rd}{(\mathbb{R},{+}\infty)}
\newcommand{\Rm}{\mathbb{R}_{-}^\ast}
\newcommand{\Rp}{\mathbb{R}_+^\ast}
\newcommand{\D}{\mathbb{D}}
\newcommand{\C}{\mathbb{C}}

\newcommand{\N}{\mathbb{N}}
\begin{document}
\date{}
\title{Incompatible Coulomb hamiltonian extensions}
\author{G. Abramovici}
\affil{Université Paris-Saclay, CNRS, Laboratoire de Physique des Solides, 
91405, Orsay, France}
\affil[*]{abramovici@lps.u-psud.fr}
\maketitle
\begin{center}
Accepted in Scientific Reports \textbf{10}, n$^\circ$~7280 (2020).
\end{center}
\begin{abstract}
We revisit the resolution of the one-dimensional Schrödinger hamiltonian with
a Coulomb $\lambda/|x|$ potential. We examine among its self-adjoint extensions
those which are compatible with physical conservation laws. In the
one-dimensional semi-infinite case, we show that they are classified on a $U(1)$
circle in the attractive case and on $\Rd$ in the repulsive one. In the
one-dimensional infinite case, we find a specific and original classification by
studying the continuity of eigenfunctions. In all cases, different extensions
are incompatible one with the other. For an actual experiment with an attractive
potential, the bound spectrum can be used to discriminate which extension is the
correct one.
\end{abstract}

\section{Introduction}

The Coulomb problem addresses the non-relativistic Schrödinger equation with a
3-dimensional Coulomb potential, restricted to one dimension; it has inspired a
vast corpus of scientific literature for the last seventy years\cite{Loudon,%
Haines,Andrews,Gesztesy,Gostev,Tsutsui,Fischer,Gordeyev,Mineev,Nunez,%
abramovici}.  Some results have been much debated. Mathematical aspects are now
fully understood, but physical ones want for more elaborated and robust
interpretation, which we provide in details here.

In this article, we study the Coulomb potential, either restricted to a
semi-infinite line, or else to a full infinite line. We will formally write the
corresponding hamiltonian $H=-d\hspace{1pt}^2\!/dx^2+V$ in dimensionless units
and $\D$ will represent the domain on which wavefunctions are defined, so the
first case corresponds to $\D=\Rp$, while the second to $\D=\R$.
When necessary, we will write $H(\D)$ instead of $H$.  One may note that the
Schrödinger equation for $\D=\Rp$ is equivalent, through a simple mapping, to
the radial one for $\D=\R^3$ in 3-dimension with zero orbital momentum, $L=0$.  

This work lies at the frontier between physics and mathematics, because Coulomb
hamiltonians $H(\Rp)$ and $H(\R)$, although defined on a physical basis, reveal
\textbf{non self-adjoint}. In such a case, one usually needs to study the
self-adjoint \textbf{extensions} $K$ of the hamiltonian. But, in this very
case, the situation is even worse, because $H$ is not even \textbf{symmetric}%
\cite{Tsutsui,abramovici} (that is, one can find two states $\varphi$ and
$\chi$ such that $\langle\varphi|H|\chi\rangle\ne
\overline{\langle\chi|H|\varphi\rangle}$).  In such a situation, one must
\textbf{restrict} the Hilbert space on which eigenstates are defined, in order
to get a symmetric operator, the self-adjoint extensions $K$ of which are
well-defined.  We call $\cal L$ this restricted Hilbert space.

When the self-adjoint extension of an operator is unique, these mathematical
manipulations are transparent because the spectral theorem applies, so the
action of the operator is defined unambiguously on any function of $\cal L$.
This is the case for almost all standard hamiltonians found in scientific
literature, which are moreover generally well defined without any restriction
(that is ${\cal L}= L^2(\D)$), so one does not need to care about all these
mathematical subtleties.  

However, $H(\Rp)$ and $H(\R)$ belong to the class of operators, which admit
\textbf{several} self-adjoint extensions. Each extension is
\textbf{incompatible} with the other, so one must \textbf{choose} only one
extension at a time, where to define a complete set of eigenstates.  From a
physical point of view, the interpretation of the operator action on a
wavefunction is ambiguous, since its definition \textbf{depends} on the
extension which is chosen.  \textit{Deficiency coefficients} are defined, which
indicate the number of degrees of freedom, for this choice. For $H(\Rp)$,
authors have found\cite{gesz3,Albeverio,Oliveira,Gitman} one continuous degree
of freedom.

\section{Motivation}

The interest of the Coulomb problem lies in its unusual properties: the fact
that hamiltonian $H(\Rp)$ and $H(\R)$ are not self-adjoint and not even
symmetric, so that one must construct maximal restrictions $\cal L$ and study
their self-adjoint extensions $K$. Our aim is to find a \textbf{physical}
interpretation of these extensions, in order to identify those which are
compatible with standard physical laws and those which are not.  

The boundary triples theory, which is proved for the Coulomb problem%
\cite{Oliveira}, establishes that any eigenfunction $\psi$ of $K$ is an
eigenfunction of $H$ with specific \textbf{boundary} conditions. This result, to
which we will refer as the \textit{boundary triples theorem}, provides a
physical interpretation of all the self-adjoint extensions to be found. We will
also benefit of all previous classifications of these extensions%
\cite{gesz3,Albeverio,Oliveira,Gitman} and repeat some of these calculations,
taking into account physical considerations.

In what concerns the semi-infinite line, all self-adjoint extensions of $H(\Rp)$
reveal compatible with physical conservation laws, so the main contribution of
this study on $H(\Rp)$ consists mainly in a more physical and pedagogical way to
construct them. However, we provide an original description of the space
parameter of these extensions, which is topologically equivalent to $U(1)$ in
the attractive case and to $(\R,\infty)$ in the repulsive one.

On the contrary, self-adjoint extensions of $H(\R)$ are not all compatible with
physical conservation laws. Indeed, their study brings a specific difficulty:
the connection of the solution defined on $\Rp$ with that defined on $\Rm$,
since the continuity of eigenfunctions at $x=0$ is not guaranteed. This has been
very debated and we propose an original connection process, which is founded on
physical conservation laws and gives \textbf{new}, although compatible, results.

Altogether, we prove a new classification of the self-adjoint extensions of
$H(\R)$, \textbf{excluding} those which are not compatible with physical
conservation laws. Accordingly, this classification maps on a space of extension
parameter, which is \textbf{reduced} compared to that of previous
classifications\cite{Gitman}, but the deficiency coefficient remains equal to 2.
The parameter space of our classification is the product of a one-dimensional
closed line by a phase similar to a gauge degree of freedom.
 
In what concerns the 3-dimension space, in spite of the mapping between its
Schrödinger equation with that of $H(\Rp)$, the corresponding classifications of
self-adjoint extensions are different (see however appendix), since the
deficiency coefficient of $H(\R^3)$ is zero\cite{Oliveira}, that is $H(\R^3)$ is
self-adjoint, when defined in $L^2(\R^3)$.

The present article is organized as follows: we will first focus on the $\D=\Rp$
case and classify all self-adjoint extensions of $H(\Rp)$, both for an
attractive potential or a repulsive one. In particular, we define and exhibit
the Dirichlet or Neumann extensions. Then, we study in details the continuation
problem in the $\D=\R$ case. Next, we study physical applications of $\D=\R^3$,
$\D=\R$ and $\D=\Rp$ cases. In a fifth part, we examine the spectral theorem.
In the next one, we exhibit the extension parameter spaces.  Finally, we will
review the highlights of this work on the Coulomb problem. Some notations and
terms are given afterwards in Tab.~\ref{tabcst}.

\section{Self-adjoint extensions in the $\Rp$ case}
\label{Rplus}

Operator $H(\Rp)$ is unbound and can not be defined on $L^2(\Rp)$, the Hilbert
space of square-integrable functions. Eigenfunctions $\phi_e$ obey equation
\begin{equation}
\label{original}
-{d^2\phi_e\over dx^2}(x)+{\lambda\over x}\phi_e(x)=e\,\phi_e(x)\quad\forall x>0
\end{equation}
where we have  multiplied Schrödinger equation by $2m/\hbar^2$, so $e$ is the
reduced energy corresponding to $E=\hbar^2 e/(2m)$; we define
$\lambda\equiv{2mqq'\over4\pi\epsilon_{\rm o}\hbar^2}$, $m$ is the mass of the
particle, $\epsilon_{\rm o}$ vacuum permittivity, $\hbar$ the reduced Planck
constant and $q$, $q'$ the electric charges.  For $e>0$ (free states of positive
energy), the solutions of (\ref{original}) read
\begin{equation}
\Psi_k(x)=\alpha_k F_\eta(k x)+\beta_k G_\eta(k x)\;,
\label{FG}\quad\hbox{with momentum }k\equiv\sqrt{e},\ 
\eta\equiv\lambda/(2k)\ \hbox{ and}
\end{equation}
\[
\!\!\!\!\!\!\!\!\!\!\!\!\!\!\!\!\!\!\!\!
F_\eta(u)\equiv C_\eta u\e^{-\ii u}M(1-\ii \eta,2,2\ii u)\;,\quad
G_\eta(u)\equiv\Re\left(2\eta{u\e^{-\ii u}\Gamma(-\ii\eta)\over C_\eta}
U(1-\ii \eta,2,2\ii u)\right),\ \hbox{with }
C_\eta\equiv\e^{-{\pi\eta\over2}}\sqrt{\pi\eta\over\sinh(\pi\eta)}.
\]
Here, $\Gamma$ is the gamma function, $M$ the regular confluent hypergeometric
function and $U$ the logarithmic confluent hypergeometric function%
\cite{Abramowitz}. Both $F_\eta$ and $G_\eta$ are continuous and
bounded, see$\!$ {\setcitestyle{numbers,open={},close={},comma}
Ref.~\cite{abramovici}} for asymptotic
behavior and other properties. The case $e=0$ extends this case when the
potential is attractive, see section~\ref{enul}.

For $e<0$ (bound states of negative energy), the solutions of (\ref{original})
read
\begin{equation}
\varphi_k(x)=\mu_k f_\eta(k x)+\nu_k g_\eta(k x)\;,
\label{fg}\quad\hbox{with momentum }
k\equiv\sqrt{-e},\ \eta\equiv\lambda/(2k)\ \hbox{ and}
\end{equation}
\[
\!\!\!\!\!\!\!\!\!
f_\eta(u)\equiv 2D_\eta u\e^{-u}U(1+\eta,2,2u)\;,\quad
g_\eta(u)\equiv 2\sqrt{|\lambda|}u\e^{-u}M(1+\eta,2,2u)\;,
\ \hbox{with }D_\eta\equiv
{|\Gamma(1+\eta)|\sqrt{|\lambda|}\over\sqrt{1-2\eta+2\eta^2\,\dig'(1+\eta)}}\ .
\]
Here, $\dig$ is the digamma function. One finds $f_\eta\in L^1(\Rp)$ $\bigcap
L^2(\Rp)\bigcap C^\infty(\Rp)$ while $g_\eta\in C^\infty(\Rp)$ and diverges as
$u\to\infty$. We have chosen $\Vert f_\eta\Vert_2=1$ in $L^2(\Rp)$.

For $\lambda<0$ so $qq'<0$ and the potential is attractive, the spectrum of any
self-adjoint extension will reveal infinite and discrete. As we shall find, all
solutions corresponding to $\eta=-n$, with $n\in\N^\ast$, belong to the same
extension and read $f_\eta(u)=-u\e^{-u}L'_n(2u)\sqrt{-2\lambda}\;n^{-3/2}$, the
standard Rydberg solution, with $L_n$ the Laguerre polynomial. They obey
Dirichlet condition $f_{-n}(0)=0$. On the other hand, for $-\eta\not\in\N^\ast$,
$f_\eta(0)\ne0$, see$\!$ {\setcitestyle{numbers,open={},close={},comma}
Ref.~\cite{abramovici}} for more
details. We will call \textit{Rydberg} states, those following $\eta=-n$ with
$n\in\N^\ast$, and \textit{non Rydberg} states the others. Note that the
definition of $g_\eta$ must be changed into
\[
g_\eta(u)\equiv2\sqrt{|\lambda|}u\e^u U(1-\eta,2,-2u)
\]
since, in that very case $\eta=-n$, $u\e^{-u}M(1-n,2,2u)$ is proportional to
$u\e^{-u}U(1-n,2,2u)$.

For $\lambda>0$ so $qq'>0$ and the potential is repulsive, the spectrum of any
self-adjoint extension will reveal discrete, with a unique bound state of
strictly negative energy, but in a specific case that we will explain further
on.

\subsection{Existence of self-adjoint extension}

The existence of self-adjoint extensions for the Coulomb potential has been
fully established in several references\cite{gesz3,Albeverio,Oliveira,Gitman}
and needs not to be discussed here again. Indeed, the deficiency coefficients
$m_\pm$ are found equal to 1, although not explicitly calculated in$\!$
{\setcitestyle{numbers,open={},close={},comma}
Ref.~\cite{gesz3}}. We will construct all self-adjoint
extensions as follows. 

We will write $H_\omega(\Rp)$ the self-adjoint extensions of $H(\Rp)$,
parametrized by $\omega$, a symbolic index, the meaning of which will be
explained later on.  The boundary triples theorem implies that $H_\omega(\Rp)$
is the restriction of $H(\Rp)$ on some domain $\cal L$ of eigenfunctions, which
we write ${\cal L}=\mathcal{D}_\omega$. We will first construct all possible
symmetric extensions of $H(\Rp)$ with different boundary conditions and find
self-adjoint ones $H_\omega(\Rp)$ as \textbf{maximal} symmetric
extensions\cite{Gitman}.

\subsection{Description of a self-adjoint extension}
\label{description}

In this part, we consider the attractive case.  Let $e_\omega<0$ be in the
spectrum of $H_\omega(\Rp)$, that is $\varphi_{k_\omega}$, with momentum
$k_\omega=\sqrt{-e_\omega}$, is an eigenfunction of $H_\omega(\Rp)$ and belongs
to $\mathcal{D}_\omega$. There is such $e_\omega$, otherwise the spectrum of
$H_\omega(\Rp)$ would be included in $\R_+$, which case we exclude later
on. $\varphi_{k_\omega}$ is proportional to $x\mapsto
f_{\eta_\omega}(k_\omega x)$ (writing $\eta_\omega=\lambda/(2k_\omega)$) because
of (\ref{fg}); indeed, $f_{\eta_\omega} \in L^2(\Rp)$, so does
$\varphi_{k_\omega}$ by definition, while $g_{\eta_\omega}$ diverges, letting
$\nu_{k_\omega}=0$. The other factor reads then
$\mu_{k_\omega}=\e^{\ii\theta_{\!\omega}}$, a constant phase factor which can be
fixed arbitrarily.

One observes that not all functions $\varphi_k$ belong to $\mathcal{D}_\omega$,
because the scalar product $\langle f_{\eta_1} |f_{\eta_2}\rangle$, which we
calculate in appendix, with arbitrary momenta
$k_i=\lambda/(2\eta_i)$, is not always zero. Let us establish this result: we
note $\geul$ the Euler constant and define function $\gb$:
\[
\gb(x)\equiv\dig(1+x)-\ln|x|-{1\over2x}+2\geul\;;
\]
then, the scalar products reads
\begin{equation}
\label{scal}
\langle f_{\eta_1}|f_{\eta_2}\rangle={D_{\eta_1}D_{\eta_2}\;\lambda^2\over
k_1^{\ 2}-k_2^{\ 2}}\times
{\gb(\eta_2)-\gb(\eta_1)\over\Gamma(1+\eta_1)\Gamma(1+\eta_2)}\ ;
\end{equation}
(this expression is valid when $\eta_1\to\eta_2$ and the limit is 1);
therefore an operator admitting all such eigenfunctions would
not be symmetric\cite{Tsutsui,abramovici}.
$\blacksquare$

Let $\mathcal{S}_{\omega}\equiv\big\{e\in\Rm\ \big/\
\langle\varphi_{k_\omega}|\varphi_{\sqrt{-e}}\rangle=0\big\}\bigcup
\big\{e_\omega\big\}$. We will prove that the set of bound states of
$H_\omega(\Rp)$ corresponds to functions generated by
 $\mathcal{B}_\omega\equiv \{\varphi_k\in L^2(\Rp)\ \big/\
{-}k^2\in\mathcal{S}_\omega\}$, so the spectrum of $H_\omega(\Rp)$ will exactly
be $\mathcal{S}_\omega\bigcup\R_+$. Let us characterize $\mathcal{S}_\omega$.
The condition $\langle\varphi_{k_1}|\varphi_{k_2}\rangle=0$ reduces to
\begin{equation}
\label{bb}
\gb(\eta_1)=\gb(\eta_2)
\end{equation}
so $\mathcal{S}_\omega=\big\{e\ \big/\ \gb\big({\lambda\over2\sqrt{-e}}\big)=
\gb(\eta_\omega)\big\}$. We study the zeros of $\gb(\eta)-\gb(\eta_\omega)$
further on. (\ref{bb}) implies that any function $\varphi_k$ orthogonal
to $\varphi_{k_\omega}$ obeys $\gb(\eta)=\gb(\eta_\omega)$ so all functions in
$\mathcal{B}_\omega$ are either proportional or orthogonal to each other. By
construction, $\mathcal{B}_\omega$ is maximal, because any function orthogonal
to $\varphi_{k_\omega}$ belongs to it; there cannot be any other eigenfunction
in $\mathcal{D}_\omega$ corresponding to a bound state , so $\{\phi_e\in
\mathcal{D}_\omega\;\big/\;e\in\mathcal{S}_\omega\}\subseteq\mathcal{B}_\omega$.
However, we cannot claim yet that this inclusion is an equality, because the
scalar product of a bound state with a free one could be different from zero.

Let us discard this possibility and thus prove  $\{\phi_e\in\mathcal{D}_\omega\;
\big/\;e\in\mathcal{S}_\omega\}=\mathcal{B}_\omega$.  Let us examine free
states.  Let $\mathcal{F}_\omega$ be the set of functions $\phi_e=\Psi_k$, with
$e>0$ and momentum $k=\sqrt{e}$, such that
\begin{equation}
\langle\varphi_{k_\omega}|\Psi_k\rangle=0\ .
\label{fb}
\end{equation}
Each $\phi_e\in\mathcal{F}_\omega$ reads $\phi_e(x)=
\alpha_\eta^\omega F_\eta(k\,x)+\beta_\eta^\omega G_\eta(k\,x)$ using
(\ref{FG}). Let us define $\gf$:
\[
\gf(x)\equiv\Re\big(\dig(1+\ii x)\big)+2\geul-\ln|x|\ ,
\]
then the scalar products $\langle f_{\eta_1}|F_{\eta_2}\rangle$ and
$\langle f_{\eta_1}|G_{\eta_2}\rangle$ calculated in appendix read
\begin{equation}
\label{scfF}
\langle f_{\eta_1}|F_{\eta_2}\rangle=\displaystyle
{D_{\eta_1}C_{\eta_2}\over4\eta_2\Gamma(1+\eta_1)}\times
{\lambda^{3/2}\over k_1^{\ 2}+k_2^{\ 2}}
\ ;\qquad
\langle f_{\eta_1}|G_{\eta_2}\rangle=\displaystyle
{D_{\eta_1}\over C_{\eta_2}}
{\gb(\eta_1)-\gf(\eta_2)\over2\Gamma(1+\eta_1)}
\times{\lambda^{3/2}\over k_1^{\ 2}+k_2^{\ 2}}\ .
\end{equation}
We define $\zeta_\eta^\omega\equiv\alpha_\eta^\omega\big/\beta_\eta^\omega$. For
$-\eta_1\not\in\N^\ast$ (non Rydberg states), using (\ref{fb}) with
(\ref{scfF}), one finds
\begin{equation}
\label{alpha}
\forall e_1=-k_1^{\ 2}\in\mathcal{S}_\omega\qquad\zeta_k^\omega=
{2\eta\over C_\eta^{\ 2}}\big(\!\gf(\eta)-\gb(\eta_1)\big)
={2\eta\over C_\eta^{\ 2}}\big(\!\gf(\eta)-\gb(\eta_\omega)\big)
\hbox{ using (\ref{bb}).}
\end{equation}

For $-\eta_1\in\N^\ast$ (Rydberg states), one finds $\langle
f_{\eta_1}|F_{\eta_2}\rangle=0$ and $\langle
f_{\eta_1}|G_{\eta_2}\rangle=(-1)^{\eta_1}{\Gamma(-\eta_1)D_{\eta_1} \over
C_{\eta_2}}{\lambda^{3/2}\over k_1^{\ 2}+k_2^{\ 2}}$ so one must choose
$\beta_{k_2}=0$ and gets $\zeta_{k_2}=\infty$.
 (\ref{alpha}) extends in this case, since $\gb(\eta_\omega)\to\infty$ when
$\eta_\omega\to-n$ with $n\in\N^\ast$.\newline (\ref{alpha}) implies that
$\Psi_k$ is orthogonal to any function $\varphi_{k_1}\in\mathcal{B}_\omega$ as
soon as it is orthogonal to $\varphi_{k_\omega}$. All free eigenfunctions of
$H_\omega(\Rp)$ must belong to $\mathcal{F}_\omega$, so they respect
(\ref{alpha}); thus, they are all orthogonal to any
$\varphi_{k_1}\in\mathcal{B}_\omega(\Rp)$; this ends our demonstration.
$\blacksquare$

Conversely, all elements in $\mathcal{F}_\omega$ are eigenfunctions of
$H_\omega(\Rp)$. In that purpose, let us establish the generalized
orthonormality of all elements in $\mathcal{F}_\omega$. Let $\phi_{e_1}$ and
$\phi_{e_2}$ be in $\mathcal{F}_\omega$, with $e_1\ne e_2$. The scalar products
$\langle F_{\eta_1}|F_{\eta_2}\rangle$, $\langle F_{\eta_1}|G_{\eta_2}\rangle$,
$\langle G_{\eta_1}|F_{\eta_2}\rangle$ and $\langle
G_{\eta_1}|G_{\eta_2}\rangle$, calculated in appendix read
\begin{equation}
\label{scFF}
\!\!\!\!\!\!\!\!\!\!\!\!\!\!\!\!\!\!\!\!
\langle G_{\eta_1}|G_{\eta_2}\rangle=\displaystyle
{\lambda\over C_{\eta_1}C_{\eta_2}}
{\gf(\eta_2)-\gf(\eta_1)\over k_1^{\ 2}-k_2^{\ 2}}+\delta(k_1-k_2)\ ;\quad
\langle F_{\eta_1}|F_{\eta_2}\rangle=\delta(k_1-k_2)\ ;\quad
\langle F_{\eta_1}|G_{\eta_2}\rangle=\displaystyle
{\lambda C_{\eta_1}\over2\eta_1C_{\eta_2}}{1\over k_1^{\ 2}-k_2^{\ 2}}\;.
\end{equation}
For $e_1\ne e_2$, we span the scalar product $\langle\phi_{e_1}|
\phi_{e_2}\rangle$ using (\ref{FG}) and (\ref{alpha}), which gives
\begin{eqnarray*}
\!\!\!\!\!\!\!\!\!\!\!\!
\langle\phi_{e_1}|\phi_{e_2}\rangle&=&
\overline{\alpha_{\eta_1}^\omega}\alpha_{\eta_2}^\omega
\langle F_{\eta_1}|F_{\eta_2}\rangle
+\overline{\alpha_{\eta_1}^\omega}\beta_{\eta_2}^\omega
\langle F_{\eta_1}|G_{\eta_2}\rangle
+\overline{\beta_{\eta_1}^\omega}\alpha_{\eta_2}^\omega
\langle G_{\eta_1}|F_{\eta_2}\rangle
+\;\overline{\beta_{\eta_1}^\omega}\beta_{\eta_2}^\omega
\langle G_{\eta_1}|G_{\eta_2}\rangle
\\
&=&{\eta_1^{\ 2}\eta_2^{\ 2}\over\lambda C_{\eta_1}C_{\eta_2}}
{\overline{\beta_{\eta_1}^\omega}\beta_{\eta_2}^\omega\over
\eta_1^{\ 2}-\eta_2^{\ 2}}\Big({0-
2\eta_1}\big(\gf(\eta_1)-\gb(\eta_\omega)\big){2\over\eta_1}
+\;2\eta_2\big(\gf(\eta_2)-\gb(\eta_\omega)\big){2\over\eta_2}
-4\big(\gf(\eta_2)-\gf(\eta_1)\big)\Big)\\
&&\qquad
+\;({|\alpha^\omega_{\eta_1}|^2+|\beta^\omega_{\eta_1}|^2})\delta(k_1-k_2)\\
&=&\delta(k_1-k_2)\ ,
\end{eqnarray*}
where $\overline{\zeta_{\eta_1}}=\zeta_{\eta_1}$ follows (\ref{alpha}).
$\blacksquare$

We have proved that all bound eigenfunctions of $H_\omega(\Rp)$ are in
$\mathcal{B}_\omega$ while all free ones are in $\mathcal{F}_\omega$.
Therefore, we get $\mathcal{D}_\omega=
\mathcal{B}_\omega\bigcup\mathcal{F}_\omega$. We
define $\widetilde{H}_\omega$ the restriction of $H(\Rp)$ on
$\mathcal{D}_\omega$. We will prove now that $\widetilde{H}_\omega$ is
symmetric, that is $\langle(\widetilde{H}_\omega\psi)|\varphi\rangle=
\langle\psi|(\widetilde{H}_\omega\varphi)\rangle$ for all
$\psi,\varphi\in\mathcal{D}_\omega$.  Let $(e_1,e_2)$ be such that
$\psi=\phi_{e_1}$ and $\varphi=\phi_{e_2}$ (depending on whether $\psi$ belongs
to the free or the bound spectrum, either $e_1\in\R^+$ or
$e_1\in\mathcal{S}_\omega$, and idem for $\varphi$ with $e_2$). One writes then
\[
\langle(\widetilde{H}_\omega\psi)|\varphi\rangle-
\langle\psi|(\widetilde{H}_\omega\varphi)\rangle
=\smallunderbrace{\overline{e_1}}_{\scriptstyle =e_1\atop}
\langle\psi|\varphi\rangle-e_2\langle\psi|\varphi\rangle
=(e_1-e_2)\langle\psi|\varphi\rangle
=0
\]
The last equality is proved by discussing whether $e_1\ne e_2$, so
$\psi=\phi_{e_1}\perp\phi_{e_2}$ $=\varphi$, following all previous discussions,
or else $e_1=e_2$.
$\blacksquare$

Let us prove that $\widetilde{H}_\omega$ is maximal ad absurdum. Since it is
symmetric, it admits a self-adjoint extension $K$, which is defined on
$\mathcal{C}\bigcup\mathcal{D}_\omega$, where $\mathcal{C}$ is some non empty
space, by hypothesis. Let us write $K_\mathcal{C}$ the restriction of $K$ on
$\mathcal{C}$. Let $\{\phi_i, i\in\mathcal{I}\}$ be a basis of
$\mathcal{D}_\omega$, and $\{\psi_j,j\in\mathcal{J}\}$ a basis of $\mathcal{C}$.
One writes
\[
K|\phi_i\rangle=\widetilde{H}_\omega|\phi_i\rangle
=\sum_{k\in\mathcal{I}} a_i^k|\phi_k\rangle\qquad
K|\psi_j\rangle=\sum_{k\in\mathcal{I}} b_j^k|\phi_k\rangle+
\sum_{l\in\mathcal{J}} c_j^l|\psi_l\rangle\ .
\]
Multiplying the first line by $\langle\psi_j|$ and the second by
$\langle\phi_i|$, one gets $b_j^i=0$, so $\langle\phi_i|\psi_j\rangle=0$
$\forall i,j$. $K$ is symmetric, so $\langle\psi_i|K|\psi_j\rangle=
\overline{\langle\psi_j|K|\psi_i\rangle}$, which implies
$c_i^j=\overline{c_j^i}$ $\forall i,j$. Eventually, we have established that
$K_\mathcal{C}$ is symmetric. From standard algebra\cite{Akhiezer}, there exists
at least an eigenfunction $\phi_0\in\mathcal{C}$, and its eigenvalue $e_0$ is
real.

Applying the boundary triples theorem, the function $\phi_0$ is a solution of
the differential equation $H(\Rp)\phi_0(x)=e_0\phi_0(x)$, with particular
boundary conditions. If $e_0<0$, one finds immediately that
$\phi_0\in\mathcal{B}_\omega$. If $e_0\ge0$, one must first write that
$\langle\varphi_k|\phi_0\rangle=0$ for all $k$ ($\varphi_k$ are the elements of
$\mathcal{B}_\omega$); following the previous construction, one eventually finds
that $\phi_0\in\mathcal{F}_\omega$. We have proved
$\phi_0\in\mathcal{D}_\omega$, which contradicts $\phi_0\in\mathcal{C}$, so the
maximality of $\widetilde{H}_\omega$ is proved.
$\blacksquare$

$\widetilde{H}_\omega$ is symmetric and maximal, that is, it is a self-adjoint
extension.  Furthermore, $\widetilde{H}_\omega$ is simple.
Concerning bound states, this results from the elimination of functions
proportional to $g_\eta$.  Concerning free states, it follows (\ref{alpha}).
Now, let $\phi_e$ be any eigenfunction included in the domain of
$H_\omega(\Rp)$.  This domain includes $\phi_{e_\omega}$, so $\phi_e$ must be
either orthogonal to $\phi_{e_\omega}$ or have eigenvalue $e_\omega$. In the
first case, $\phi_e$ belongs to $\mathcal{D}_\omega$. In the second case, it is
proportional to $\phi_{e_\omega}$ (still resulting from the elimination of
functions proportional to $g_{\eta_\omega}$).  This proves that the domain of
$H_\omega(\Rp)$ is included in that of $\widetilde{H}_\omega$, so
$\widetilde{H}_\omega$ is an extension of $H_\omega(\Rp)$. Self-adjoint
extensions are maximal, so $\widetilde{H}_\omega=H_\omega(\Rp)$, which is
therefore completely determinate.
$\blacksquare$

\begin{figure}[t]
\begin{center}
\includegraphics[width=9cm]{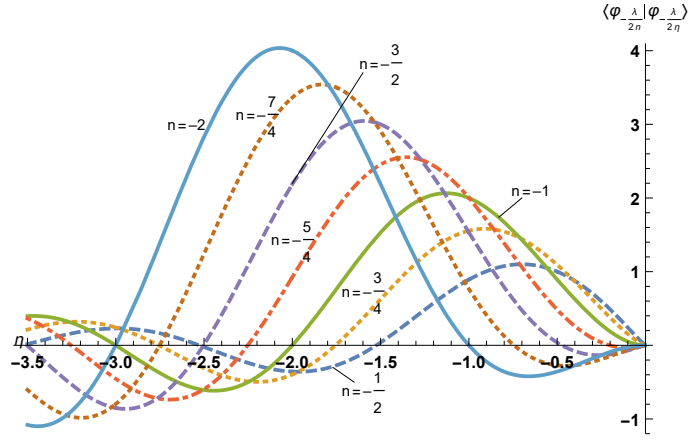}
\end{center}
\caption{Here are the curves $\eta\mapsto
\langle\varphi_{-{\lambda\over2n}}|\varphi_{-{\frac\lambda{2\eta}}}\rangle$, for
$n=-1/2$ (dashed line), $n=-3/4$ (dotted line), $n=-1$ (plain line), $n=-5/4$
(dot-dashed line), $n=-3/2$ (dashed line), $n=-7/4$ (dotted line) and $n=-2$
(plain line). The zeros of each curve read $\eta={\lambda\over2\sqrt{-e}}$ where
$e\in\mathcal{S}_\omega$, with $\omega=-\gb(n)$, as explained further on.  The
curves seem to form pairs corresponding to $(n,n+1)$, in particular, one could
believe that each pair intersects on the $\eta$-axis (abscissa), but this is
wrong, except for $(n,n+1)=(-2,-1)$ which correspond to the same Rydberg set
$\mathcal{S}_\infty$. All the other intersections are only close to zero, so
that, indeed, $\mathcal{S}_{-\gb(n)}\ne\mathcal{S}_{-\gb(n+1)}$.  $\eta=n$ is
missing, because
$\langle\varphi_{-{\lambda\over2n}}|\varphi_{-{\lambda\over2n}}\rangle\ne0$.}
\label{spectra}
\end{figure}

\subsection{Classification in the attractive case}
\label{attractive}

Set $\mathcal{S}_\omega$ contains the zeros of $\eta\mapsto
\gb\big({-}\big({\lambda\over2\eta}\big)^2\big)-\gb(e_\omega)$, which we
represent for several values of $\omega$ in Fig.~\ref{spectra}.  To characterize
each set $\mathcal{B}_\omega$, we follow the results in$\!$
{\setcitestyle{numbers,open={},close={},comma} Ref.~\cite{Oliveira}} and define
$\widetilde{\mathcal{B}}_\omega=\big\{\varphi_k\in L^2(\Rp)\ \big/\
{\partial\varphi_k(x)\over|\lambda|\partial x}+\varphi_k(x)\ln(|\lambda|x)$ $=
\omega\varphi_k(x)\big\}$.  This condition differs from the more usual one
${\partial\phi(k\,x)\over\partial x}=\omega\phi(k\,x)$. Another possible
characterization is given in$\!$
{\setcitestyle{numbers,open={},close={},comma} Ref.~\cite{gesz3}}.  For a
given number $\omega\in\R$, we define $\eta_\omega$ to be any solution of
$\gb(\eta)=-\omega$ (one can chose the highest $\eta$, as we will prove further
on that this set has a maximum).

Let us prove $\mathcal{B}_\omega=\widetilde{\mathcal{B}}_\omega$. First, we will
show that two functions in $\widetilde{\mathcal{B}}_\omega$ are either
proportional or orthogonal. One finds, for non Rydberg eigenfunctions
($-\eta\not\in\N^\ast$),
\begin{eqnarray*}
{\partial f_\eta\big({\lambda x\over2\eta}\big)\over\partial x}
={\lambda\over2\eta}f'_\eta\Big({\lambda x\over2\eta}\Big)\quad
&\hbox{so}&\lim_{x\to0}
{\partial f_\eta\big({\lambda x\over2\eta}\big)\over|\lambda|\partial x}
+f_\eta\Big({\lambda x\over2\eta}\Big)\ln(|\lambda|x)
=-{\gb(\eta)D_\eta\over\Gamma(1+\eta)}\\
\hbox{while }\ 
\lim_{x\to0}f_\eta\Big({\lambda x\over2\eta}\Big)={D_\eta\over\Gamma(1+\eta)}
\qquad &\hbox{so}&\lim_{x\to0}
{{\partial f_\eta({\lambda x\over2\eta})\over|\lambda|\partial x}
+f_\eta\big({\lambda x\over2\eta}\big)\ln(|\lambda|x)\over 
f_\eta\big({\lambda x\over2\eta}\big)}=-\gb(\eta)\ .
\end{eqnarray*}
Thus, it comes that all elements in $\widetilde{\mathcal{B}}_\omega$ verify
$\omega=-\gb(\eta)$, so, using (\ref{bb}), the proposition is proved, except for
Rydberg states such that $-\eta\in\N^\ast$. For these, the last limit gives
$\infty$. However, these eigenfunctions are well known and indeed orthogonal
(see section~\ref{dirichlet}), so the result extends to this case immediately.
Conversely, any index $\eta$ corresponding to $\varphi_k\in \mathcal{B}_\omega$
verifies $\gb(\eta)=\gb(\eta_\omega)=-\omega$.  Eventually, this proves
$\widetilde{\mathcal{B}}_\omega= \mathcal{B}_\omega$.
$\blacksquare$

Let's define $\widetilde{\mathcal{F}}_\omega=
\big\{\Psi_k\in L^\infty(\Rp)\ 
\big/\ {\partial\Psi_k(x)\over|\lambda|\partial x}+\Psi_k(x)\ln(|\lambda|x)$ 
$=\omega\Psi_k(x)\big\}$.
We will prove now that $\widetilde{\mathcal{F}}_\omega=\mathcal{F}_\omega$.
We first show $\mathcal{F}\subset\widetilde{\mathcal{F}}_\omega$. One finds
\begin{eqnarray*}
\lim_{x\to0}
\lim_{x\to0}F_\eta\Big({\lambda x\over2\eta}\Big)=0\;;&&
\lim_{x\to0}F_\eta\Big({\lambda x\over2\eta}\Big)\ln(|\lambda|x)=0\;;\quad
{\partial F_\eta\big({\lambda x\over2\eta}\big)\over|\lambda|\partial x}
={C_\eta\over2\eta}\\\hbox{then}\quad
\lim_{x\to0}G_\eta\Big({\lambda x\over2\eta}\Big)={1\over C_\eta}&&
\hbox{and }\ 
\lim_{x\to0}
{\partial G_\eta\big({\lambda x\over2\eta}\big)\over|\lambda|\partial x}+
G_\eta\Big({\lambda x\over2\eta}\Big)\ln(|\lambda|x)=-{\gf(\eta)\over
C_\eta}\;;
\end{eqnarray*}
so, considering any $\phi_e(x)=\alpha_k^\omega F_\eta(k\,x)+
\beta_\eta^\omega G_\eta(k\,x)\in\mathcal{F}_\omega$ with
$\beta_k^\omega\ne0$, one gets
\begin{eqnarray*}
\lim_{x\to0}
{\alpha_k^\omega{\partial F_\eta({\lambda x\over2\eta})\over|\lambda|\partial x}
+\beta_\eta^\omega
{\partial G_\eta({\lambda x\over2\eta})\over|\lambda|\partial x}\over
\alpha_k^\omega F_\eta\big({\lambda x\over2\eta}\big)
+\beta_\eta^\omega G_\eta\big({\lambda x\over2\eta}\big)}+\ln(|\lambda|x)
&=&
\zeta_k^\omega\lim_{x\to0}{
{\partial F_\eta({\lambda x\over2\eta})\over|\lambda|\partial x}\over
G_\eta\big({\lambda x\over2\eta}\big)}
+\lim_{x\to0}{
{\partial G_\eta({\lambda x\over2\eta})\over|\lambda|\partial x}\over
G_\eta\big({\lambda x\over2\eta}\big)}+\ln(|\lambda|x)\\
&=&2\eta(\gf(\eta)-\gb(\eta_\omega)){1\over2\eta}-\gf(\eta)
=-\gb(\eta_\omega)\ .
\end{eqnarray*}
while, for $\beta_k^\omega=0$, which corresponds to Rydberg states, one gets
\[
\lim_{x\to0}{
{\partial F_\eta({\lambda x\over2\eta})\over|\lambda|\partial x}
+F_\eta\big({\lambda x\over2\eta}\big)\ln(|\lambda|x)\over
F_\eta\big({\lambda x\over2\eta}\big)}
=\infty\ .
\]
This proves exactly that $\phi_e$ belongs to $\widetilde{\mathcal{F}}_\omega$.
Reversely, let us show that any element $\phi_e\in
\widetilde{\mathcal{F}}_\omega$ belongs to $\mathcal{F}_\omega$. Using
(\ref{FG}), one writes $\phi_e=\alpha_kF_\eta+\beta_kG_\eta$. Then, from the
definition of $\widetilde{\mathcal{F}}_\omega$, one gets
\[
\alpha_k{C_\eta\over2\eta}-\beta_k{\gf(\eta)\over C\eta}
=\beta_k{\omega\over C_\eta}
\iff\left\{\begin{array}{lrl}
\hbox{if }\beta_k\ne0&\zeta_k\,=&
{\scriptstyle2\eta\scriptstyle\over C_\eta^{\ 2}}
\big(\omega+\gf(\eta)\big)\;;\\
\hbox{if }\beta_k=0&\zeta_k\,=&\infty\;;
\end{array}\right.
\]
and the scalar product $\langle\varphi_{\eta_\omega}|\phi_e\rangle$ reads
\[
\langle\varphi_{\eta_\omega}|\phi_e\rangle=
\alpha_k\langle\varphi_{\eta_\omega}|F_\eta\rangle
+\beta_k\langle\varphi_{\eta_\omega}|G_\eta\rangle
={\mu_{k_\omega}\lambda^{3/2}\over2(k_\omega^{\ 2}+k^2)\Gamma(1+\eta_\omega)}
\Big({\alpha_kC_\eta\over2\eta}+{\beta_k\over C_\eta}(-\omega-\gf(\eta)\Big)
=0
\]
so $\phi_e\in\mathcal{F}_\omega$. The case $\beta_k=0$ corresponds to the
Rydberg one, $\zeta_k=\infty=\omega$, and $|\varphi_{\eta_\infty}\rangle$ is
orthogonal to all states in $\mathcal{F}_\infty$.
$\blacksquare$

Our classification is coherent with that of$\!$
{\setcitestyle{numbers,open={},close={},comma}
Ref.~\cite{Oliveira}}, all self-adjoint extensions of
$H(\Rp)$ are classified by $\omega\in\R$. The topology of the parameter space is
studied in section~\ref{U1struct}.

\subsection{Classification in the repulsive case}

We now consider the repulsive case. The physical situation is very different to
the previous one, for instance, one observes that there is no Rydberg state,
that is no eigenfunction obeying $\phi_e(0)=0$, however many steps of the
calculations are similar, so we will only point out the differences.

\begin{figure}[H]
\begin{center}
\includegraphics[width=7cm]{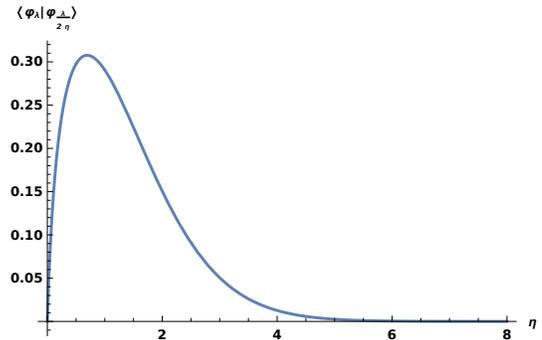}
\end{center}
\caption{$\langle\varphi_\lambda|\varphi_{\lambda\over2\eta}\rangle$ versus
$\eta$ in the repulsive case; the choice $\eta_1=\frac12$ is arbitrary, curves
obtained for other values are similar.\penalty-10000}
\label{scul}
\end{figure}

\begin{figure}[H]
\begin{center}
\includegraphics[width=7cm]{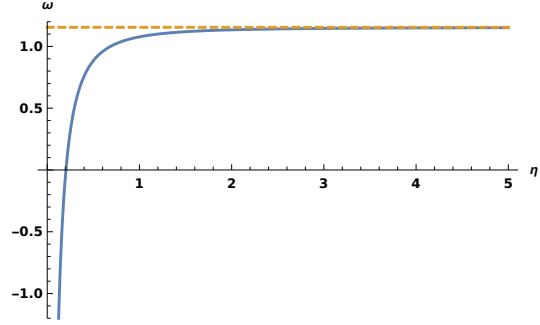}
\end{center}
\caption{$\omega$ versus $\eta$ in the repulsive case. The asymptote
$\omega=2\geul$ is drawn with a dashed line.}
\label{omegaR}
\end{figure}

Keeping the definition of $\gb$ with $\eta>0$, one finds (\ref{scal}) with the
opposite sign. Then, (\ref{bb}) has no solutions, but the existence of a bound
state will hold in the repulsive case,  which means that it is a unique bound
state. This is true for all $\omega$, see for instance Fig.~\ref{scul}, and
confirmed by the bijectivity of $\eta\mapsto\omega(\eta)$, as one observes on
Fig.~\ref{omegaR}.  However, (\ref{alpha}) extends in the repulsive case, where
$\eta_\omega$ stands for the unique bound state in $H_\omega(\Rp)$ and the sign
is also changed.
\begin{multicols}{2}
In the free spectrum, a similar sign difference occurs: 
\[
\langle G_{\eta_1}|G_{\eta_2}\rangle=
{\lambda\over C_{\eta_1}C_{\eta_2}}
{\gf(\eta_1)-\gf(\eta_2)\over k_1^{\ 2}-k_2^{\ 2}}+\delta(k_1-k_2)\ ,
\]
where the definition of $\gf$ is unchanged. The scalar product expression
$\langle G_{\eta_1}|F_{\eta_2}\rangle$ is unchanged but mind that its
real sign is also changed after that of $\eta$. Eventually, the demonstration
that all functions in $\mathcal{F}_\omega$ respect
$\langle\Phi_{e_1}|\Phi_{e_2}\rangle$=0 holds, and, consequently, the
determination of $H_\omega(\Rp)$ is formally identical.

The characterization of $\mathcal{F}_\omega$ is performed with index
\[
\omega(\eta)=\lim_{x\to0}{
{\partial\phi_e({\lambda x\over2\eta})\over\lambda\partial x}
-\phi_e\big({\lambda x\over2\eta}\big)\ln(\lambda x)\over
\phi_e\big({\lambda x\over2\eta}\big)}
\]

\end{multicols}
(note the sign difference). With this new definition, index $\omega(\eta)$
has the same expression than in the attractive case. The demonstration is
straight forward for the bound states; for free ones, one finds
\[
\lim\limits_{x\to0}
{\partial G_\eta\big({\lambda x\over2\eta}\big)\over\lambda\partial x}
-G_\eta\Big({\lambda x\over2\eta}\Big)\ln(\lambda x)={\gf(\eta)\over C_\eta}
\quad\hbox{and}\quad
\lim\limits_{x\to0}G_\eta\Big({\lambda x\over2\eta}\Big)=-{1\over C_\eta}\ ;
\]
the expression obtained for $F_\eta$ are unchanged, but mind that the real sign
is changed after that of $\eta$. Eventually, there is no sign change for index
$\omega(\eta)$ in all cases. We plot this function in Fig.~\ref{omegaR} and
observe another major difference: it maps $\Rp$ on $]-\infty,2\geul]$. 
\noindent
As a consequence, $\omega$ is bounded from above. The particular value
$\omega=2\geul$ brings a very peculiar situation and must be studied elsewhere.

\subsection{Existence of a bound state}
\label{existence}

The classification of self-adjoint extensions of $H(\Rp)$ is achieved, except
that we did not prove the existence of a bound eigenstate
$|\varphi_{k_\omega}\rangle$ of $H_\omega(\Rp)$ associated to the eigenvalue
$e_\omega\le0$ in both attractive and repulsive cases. 

We suppose  ad absurdum that the spectrum is included in $\R_+$. We consider two
eigenfunctions $\Psi_{k_1}$ and $\Psi_{k_2}$. We can choose momenta $k_1\ne
k_2$, otherwise $H_\omega(\Rp)$ would only act on functions $F_{\eta_1}$ and
$G_{\eta_1}$, which norm are infinite; no integrable function could be
constructed and this extension would not be physical. The same argument holds
if there is only one eigenfunction.

Using (\ref{FG}) and (\ref{scFF}), one gets
\[
\langle\Psi_{k_1}|\Psi_{k_2}\rangle={\lambda\over k_1^{\ 2}-k_2^{\ 2}}
\Big(-{C_{\eta_1}\overline{\alpha_{k_1}}\beta_{k_2}\over2\eta_1C_{\eta_2}}+
{C_{\eta_2}\alpha_{k_2}\overline{\beta_{k_1}}\over2\eta_2C_{\eta_1}}
+{\overline{\beta_{k_1}}\beta_{k_2}\over C_{\eta_1}C_{\eta_2}}
\big(\gf(\eta_1)-\gf(\eta_2)\big)\Big)
=0\ .
\]
If $\beta_{k_1}=0$ and $\beta_{k_2}\ne0$, one gets $\alpha_{k_1}=0$, which is
impossible since $\Psi_{k_1}\ne0$. So, either both $\beta_{k_i}$ are zero, or
both are different from zero. In the first case, this property extends to all
free states, which are therefore all Rydberg free ones; thus, $H_\omega(\Rp)$
can extend on all standard Rydberg solutions, including bound ones, which
contradicts our hypothesis.

The remaining case leads to $\beta_{k_i}\ne0$ $\forall i=1,2$, which means that
momenta $k_i$ correspond to non Rydberg states. Multiplying by
$C_{\eta_1}C_{\eta_2}\over\overline{\beta_{k_1}}\beta_{k_2}$, one gets
\begin{equation}
\label{boundstate}
-{C_{\eta_1}^{\ \ 2}\,\overline{\zeta_{k_1}}\over2\eta_1}
+{C_{\eta_2}^{\ \ 2}\,\zeta_{k_2}\over2\eta_2}
+\gf(\eta_1)-\gf(\eta_2)=0\ .
\end{equation}

One can assume $\beta_{k_i}$ real, without loss of generality. Let us define the
real and purely imaginary parts of eigenstates $\Psi_{k_i}$, $\Psi_{k_i}^{\rm r}
=\Re(\Psi_{k_i})$ and $\Psi_{k_i}^{\rm i}=\Im(\Psi_{k_i})$. Since
(\ref{original}) is real, both $\Psi_{k_i}^{\rm r}$ and $\Psi_{k_i}^{\rm i}$ are
eigenfunctions associated to the same momentum $k_i$. By construction
($\beta_{k_i}$ real), $\Psi_{k_i}^{\rm i}\;\propto\;F_{\eta_i}$, which
corresponds to a Rydberg state (because $\gb(\frac\lambda{2k})=\infty$,
cf.~\ref{attractive}, in which this item holds both for repulsive or attractive
case) and is contradictory, unless $\Psi_{k_i}^{\rm i}=0$. Altogether, this
implies that $\zeta_{k_i}$ is real $\forall i=1,2$. Eventually, one gets
\begin{equation}
\label{zeta}
{C_{\eta_1}^{\ \ 2}\,\zeta_{k_1}\over2\eta_1}-\gf(\eta_1)=
{C_{\eta_2}^{\ \ 2}\,\zeta_{k_2}\over2\eta_2}-\gf(\eta_2)\ ,
\end{equation}
so ${C_\eta^{\ 2}\zeta_k\over2\eta}-\gf(\eta)$ is a real constant, which we
write $\widetilde{\omega}$.  From the classifications above, one observes that
all functions $\Psi_k$ are eigenfunctions of $H_{\widetilde{\omega}}(\Rp)$,
which proves an extension of $H_\omega(\Rp)$ and therefore contains bound
eigenstates. We have reached a contradiction. In all cases, we have shown that
there is at least one bound state.
$\blacksquare$

In the repulsive case, it is the only one. In the attractive case, they are
infinitely many; let us study that of highest energy.

\subsection{Maximum of $\mathcal{S}_\omega$}

\begin{figure}[H]
\begin{center}
\includegraphics[width=7cm]{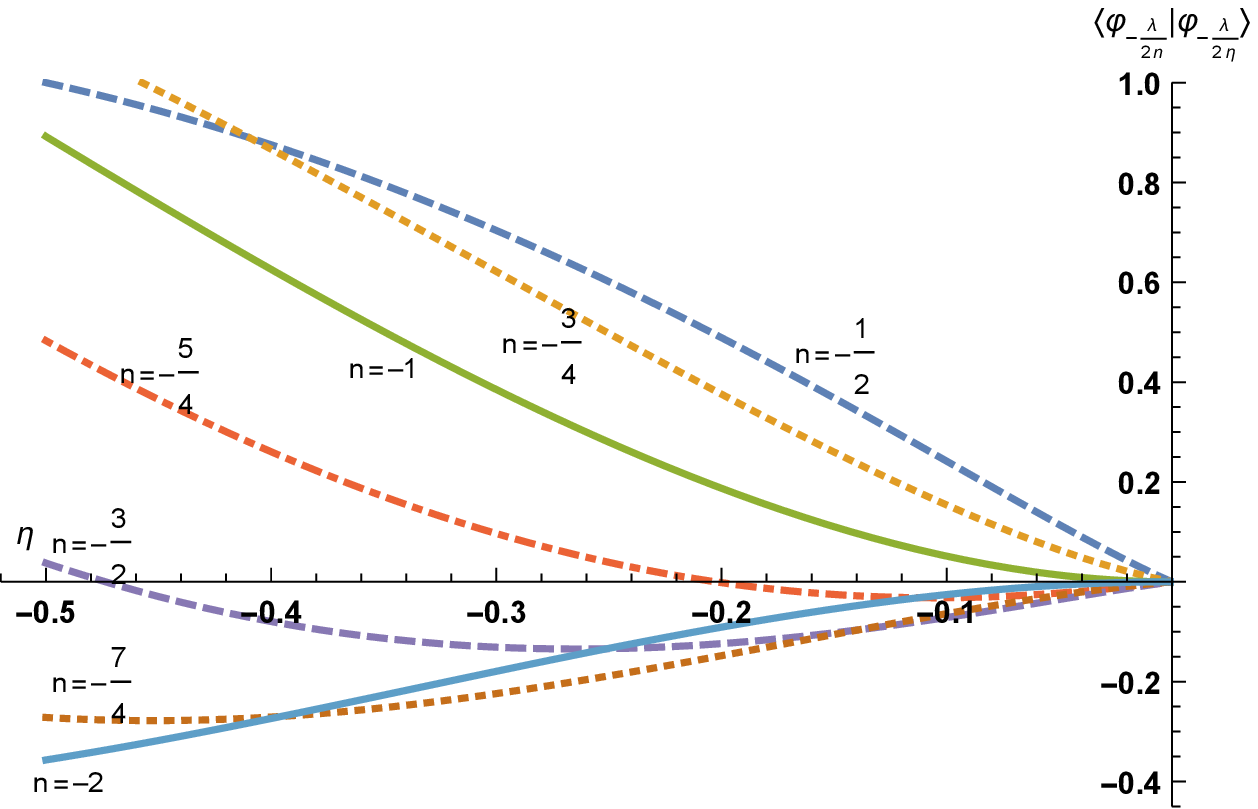}
\end{center}
\caption{Here is a zoom of Fig.~\ref{spectra} in the interval
$[-{1\over2},0]$.}
\label{spectrG}
\end{figure}

For the attractive case, $\eta<0$, so one is interested in the maximal value
$\eta_{\omega\;\rm max}$ corresponding to the maximum of $\mathcal{S}_\omega$.
There exists such a maximum, this is visible on Fig.~\ref{spectrG}, which is a
close focus of Fig.~\ref{spectra} in the interval $[-{1\over2},0]$.
\noindent
To be more precise, the slope of curve $\eta\mapsto
\langle\varphi_{-{\lambda\over2n}}|\varphi_{-{\lambda\over2\eta}}\rangle$ at
$\eta=0$ reads $2\over\Gamma(n)$ which indicates that the curves corresponding
to Rydberg eigenstates, $-n\in\N^\ast$, are flat, while the sign of the slope of
the other curves is positive for $[n]_+$ even and negative for $[n]_+$ odd.
Therefore, the maximal $\eta<0$, related to an energy $e\in\mathcal{S}_\omega$
is the first zero from the right. The only difficult case would be that of the
flat curves; these however correspond to the standard Rydberg solutions
$n\in-\N^\ast$, the maximal value of which is indeed $-1$.
$\blacksquare$

\subsection{Infinite energy state}

In what precedes we exclude value $\eta=0$. Limit $\eta\to0$ of
eigenfunctions corresponding to bound states reads $f_0(u)=\e^{-u}$, but using
rescaled $\varphi_k(x)=f_\eta(kx)$ and renormalizing by $D_\eta$, one gets
\[
\lim_{\eta\to0^\pm}\left(f_\eta\Big({\lambda x\over2\eta}\Big)\right)=0
\]
for all $x\in\Rp$ but not for $x=0$ ($\pm=+$ in the repulsive case, $\pm=-$ in
the attractive one). The so called infinite energy $-\infty$ would correspond to
a singular distribution with $\{0\}$ support. Looking for such a solution, one
substitutes $\varphi_0=\sum_{n=0}^\infty a_n\delta^{(n)}$ in (\ref{original}).
In the $\eta=0$ limit, all coefficients $a_n$ are found zero, which definitely
discards such solution.

The limit $\eta\to0$ of eigenfunctions corresponding to free states reads
$F_0(x)=\sin(x)$ and $G_0(x)=\cos(x)$.  Using rescaled
$\Phi_0(x)=\alpha_0F_0(kx)+\beta_0G_0(kx)$ (but no renormalization is needed,
since the limit of $C_\eta$ is 1), one gets
\[
\lim_{\eta\to\infty}F_{\pm\eta}\Big({\lambda x\over2\eta}\Big)=0\quad\hbox{and}
\quad\lim_{\eta\to\infty}G_{\pm\eta}\Big({\lambda x\over2\eta}\Big)=1\ .
\]
The first is zero so the limit of eigenfunctions when $e\to+\infty$ is the
constant function $\Psi(x)=1$.

Eventually, we should compare these limits to the solutions of (\ref{original}),
where $\eta$ is replaced by~0. They read $\phi_\infty(x)=a x+b$, but $a\ne0$
gives divergent non physical functions, so, up to an arbitrary phase, one finds
$b=1$, which is the $e=+\infty$ limit.
$\blacksquare$

Incidentally, we are in position to discuss the long-standing claim\cite{Loudon}
of a solution $|\phi_{-\infty}\rangle$ with energy $-\infty$: we see that this
solution does not exist, putting an end to this old story.

\subsection{Discussion of some particular cases}

\subsubsection{Dirichlet solutions}
\label{dirichlet}

We consider the attractive case. When $\omega\to\pm\infty$, one gets the
Dirichlet condition $\phi_e(0)=0$. For bound states, this can be shown by
examining the limit $\varphi_k(0^+)=D_\eta/\Gamma(1+\eta)$, which we give in
section~\ref{attractive} and which is also valid in the repulsive case. For
free states, this follows, firstly, from the fact that $\zeta_k\to\infty$, as
shown in the same section, which implies $\beta_k\to0$ so
$\phi_e\;\propto\;F_\eta$, secondly from the limit $F_\eta(0^+) =0$, still
proved in that section. Then, the corresponding values of $\mathcal{S}_\infty$ are
exactly $-\lambda^2/(4n^2)$, for all $n\in\N^\ast$, which is the standard
Rydberg spectrum (in dimensionless unit). Moreover, the function
$\eta\mapsto\omega(\eta)=-\gb(\eta)$ respects $\omega(\eta+1)=\omega(\eta)$ for
all $\eta=-n$ with $n\in\N^\ast$ and only for these values.

In the repulsive case, one must recall that there is no Rydberg state, even in
the limit $\omega\to-\infty$, so this discussion is not relevant for this case.

\subsubsection{Neumann solutions}

The case $\omega=0$ will be called the Neumann solutions, because the finite
part\cite{Hadamard} of $\phi'_e\in\mathcal{D}_0$, where the essential divergent
function $\ln(k\,x)$ is left aside, is exactly zero at $x=0$. These functions
are very close to the anomalous solutions of$\!$
{\setcitestyle{numbers,open={},close={},comma}
Ref.~\cite{abramovici}}, however those do not belong to a
single extension: they are proportional to $G_\eta$ in the free spectrum and
correspond to $\zeta_k=0$. We have shown previously that $\zeta_k= {2\eta\over
C_\eta^{\ 2}}(\gf(\eta)-\gb(\eta))$, which zeros are not exactly periodic, on
the contrary, each one belongs to a different extension.  The very small
difference between any such anomalous state and the closest Neumann one explains
the small violation of orthogonality that was calculated\cite{abramovici} (when
$\eta\to\infty$, the difference between Neumann and anomalous solutions tends to
zero, as well as the scalar products between anomalous solutions).

As is well understood now, the correct choice is to consider functions in
$\mathcal{B}_0$. On the contrary, it is not physical to consider any two
anomalous states together\cite{nunez2}, because they do not belong to the same
self-adjoint extension. 

\subsubsection{Physical interpretation of $\omega$}

We did not give any physical interpretation of $\omega$ yet. It is the limit of
the ratio ${\partial\phi(x)\over|\lambda|\partial x}/\phi(x)$ between the
derivative of the wavefunction and the wavefunction itself when $x\to0$,
\textbf{after subtracting the divergent term} $\pm\ln(|\lambda|x)$ ($\pm=+$ when
the potential is attractive, $\pm=-$ when it is repulsive).

This ratio relates to the initial condition that one fixes at $x=0$ when solving
Schrödinger equation $H\phi=E\phi$. An infinite ratio corresponds to choosing
Dirichlet conditions, a zero ratio to Neumann ones, and any finite value
in-between means fixing an intermediate condition, that mixes $\phi$ and
$\phi'$.

\subsubsection{Solutions of zero energy}
\label{enul}

Writing $\R_+$, we have indicated that $0$ must be included in the free
spectrum.  This is worth giving some details.

The solutions of (\ref{original}) for $e=0$ and $\lambda<0$ read
\[
\Psi_0(x)=\alpha j(x)+\beta y(x)\,;\quad
j(x)=\sqrt{\phantom{aaa}}\hspace{-1.67em}|\lambda|x
J_1\!\big(2\sqrt{\phantom{aaa}}\hspace{-1.67em}|\lambda|x\big)\,;\quad
y(x)=\sqrt{\phantom{aaa}}\hspace{-1.67em}|\lambda|x\,
Y_1\!\big(2\sqrt{\phantom{aaa}}\hspace{-1.67em}|\lambda|x\big),
\]
where $J_1$ and $Y_1$ are Bessel functions of, respectively, the first and
second kind. That for $\lambda>0$ read
\[
\Psi_0(x)=\alpha\,\iota(x)+\beta\,\kappa(x)\,;\quad
\iota(x)=\sqrt{\phantom{aaa}}\hspace{-1.68em}|\lambda|x
I_1\!\big(2\sqrt{\phantom{aaa}}\hspace{-1.68em}|\lambda|x\big)\,;\quad
\kappa(x)=\sqrt{\phantom{aaa}}\hspace{-1.68em}|\lambda|x
K_1\!\big(2\sqrt{\phantom{aaa}}\hspace{-1.68em}|\lambda|x\big),
\]
where $I_1$ and $K_1$ are modified Bessel functions of, respectively, the first
and second kind.

We have extended the notations we use for free states, because these solutions
are indeed the limit of those ones, $j\;\propto\;F_{-\infty}$, $y\;\propto\;
G_{-\infty}$, $\iota\;\propto\;F_\infty$ and $\kappa\;\propto\;G_\infty$. The
attractive case $\eta<0$ brings nothing special, solutions $j$ and $y$ have the
standard properties of the eigenfunctions corresponding to free states ; one may
say that this limit is regular.

\begin{figure}[H]
\begin{center}
\includegraphics[width=7cm]{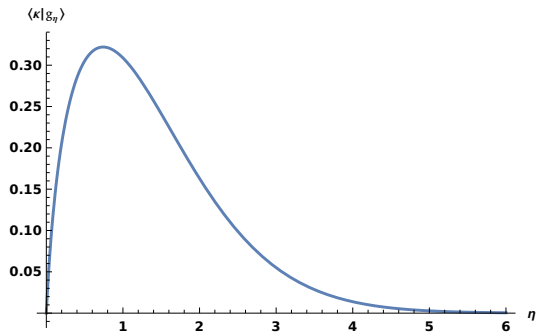}
\end{center}
\caption{$\langle\kappa|g_\eta\rangle$ versus $\eta$.}
\label{Ginfty}
\end{figure}

On the contrary, the repulsive case $\eta>0$ is extraordinary. Instead of heavy
mathematical considerations, let us explain the situation by hand. When one
looks at the curves of functions $x\mapsto F_\eta(x)$ and $x\mapsto G_\eta(x)$,
for increasing $\eta$, one observes that there are two regions $x\in[0,x_\eta]$
and $x\in[x_\eta,\infty[$, where $x_\eta$ is a separating parameter which we do
not care to define properly here. In region $[0,x_\eta]$, $F_\eta$ resembles
eigenfunction $g_\eta$ (in other words, it grows considerably, as if it
were diverging) and $G_\eta$ resembles eigenfunction $f_\eta$ (in other
words, it becomes exponentially small).  But, as these functions reach $x_\eta$,
they rapidly change shape and behave like those corresponding to standard free
states (bounded and oscillating).
\begin{multicols}{2}
This peculiar behavior, resembling bound states in a first region then free ones
afterwards, reaches its climax when $\eta\to\infty$, where $x_\eta\to\infty$:
indeed, solution $\iota$ is diverging, while $\kappa\in L^1(\Rp) \bigcap
L^2(\Rp)$.  In this very case, $F_\infty$ must be discarded and the scalar
products between $G_\infty$ and eigenfunctions $f_\eta$ reads
\[
\langle\kappa|f_\eta\rangle={2\eta D_\eta\over\Gamma(1+\eta)}
\Big(1+2\eta\big(\ln(\eta)-\Gamma(1+\eta)\big)\Big)
\]
and is non zero, as observed on Fig.~\ref{Ginfty}.
The orthogonal combination of eigenfunctions $F_\eta$ and $G_\eta$ is governed
by ratio
\[
{\alpha_\eta^\infty\over\beta_\eta^\infty}=
{2\eta\over C_\eta^{\ 2}}\Big(\Re\big(\Gamma(1-\ii\eta)\big)-\ln(\eta)\Big)\ .
\]

\end{multicols}
Our guess is that, in the repulsive case, a singular contribution $\delta(E)$
appears in the density of states, contrary to the situation of the attractive
case. This belief is founded by the existence of a bound eigenstate, to which
corresponds an integrable function, with eigenvalue $e=0$.

Eventually, one is interested in the corresponding value of index
$\omega(\infty)$. One finds $\omega(\infty)=2\geul$. Moreover, the limit
of regular bound eigenfunction $\varphi_k$, when $\eta\to\infty$, does not
exist, so there is exactly one bound eigenstate of energy $e=0$ corresponding to
$\omega(\infty)=2\geul$, which is exactly that proportional to $\kappa$. 

\section{The real line problem}

We discuss here the attractive case for $\D=\R$. We should point out that there
was no need to use of any physical constraint in the previous cases, except when
we have discarded the hypothesis of a unique energy $e>0$ or that with only two
energies $e_1>e_2>0$.  On the contrary, our determination of self-adjoint
extensions for $\D=\R$ is much more involved with physical laws.  Our aim is to
classify self-adjoint extensions that are compatible with physical constraints.

We note $\phi_e$ eigenfunctions defined on $\D$, $\phi_e^>$ their restriction on
$\Rp$ and $\phi_e^<$ that on $\Rm$. $\phi_e^>$ obeys (\ref{original}), while
$\phi_e^<$ obeys
\begin{equation}
\label{negatif}
-{\partial^2\phi_e\over\partial x^2}(x)-{\lambda\over x}\phi_e(x)=
e\phi_{e}(x)\quad \forall x<0\ .
\end{equation}
The continuity of all functions $\phi_e$ as well as their derivatives is easily
verified for all $x\ne0$ from (\ref{original}) and (\ref{negatif}).
The only difficulty lies at $x=0$. Let us define the self-adjoint extensions of
$H(\R)$.

\subsection{Self-adjoint extensions}

The mathematical classification of all self-adjoint extensions, for $\D=\R$, has
already been done\cite{Oliveira} but no effort has been made yet to interpret 
these from a physical point of view. We want to select, among all extensions,
only those, the eigenfunctions of which describe physical states.

Usually, authors impose continuous boundary conditions for all wavefunctions and
their derivative\cite{Hill,Mitchell,Lent,Mukherji} but these conditions reveal
often too restrictive and other boundary conditions have been suggested%
\cite{Bastard,Cardona}. So, we choose weaker and universal constraints, which
are compatible with any of these conditions and fit with all experimental
observations: the density of probability cannot vary discontinuously, therefore
$\rho=|\phi|^2$ must be continuous. $\rho$ also obeys the conservation of
probability law (\ref{conservj}). This implies eventually that $d{\bf j}/dx$ be
defined at all $x\in\R$. 

We introduce boundary condition $\mathcal{C}(\theta)$:
\[
\left.
\begin{array}{c}
\displaystyle
\lim_{\varepsilon\to0}\Big(\phi(\varepsilon)
-\e^{\ii\theta}\phi(-\varepsilon)\Big)=0\ ;\\
\displaystyle
\lim_{\varepsilon\to0}\Big(\phi'(\varepsilon)\phi(-\varepsilon)
-\e^{\ii\theta}\phi'(-\varepsilon)\phi(\varepsilon)\Big)=0\ ;
\end{array}\right\}\quad\theta\in[0,2\pi[\;;
\]
we will find that physical states do respect conditions $\mathcal{C}(\theta)$.
We will therefore construct self-adjoint extensions, with these boundary
conditions. More precisely, we will show that there are at maximum two values
$\theta_1$ and $\theta_2$, such that eigenfunctions obey
$\mathcal{C}(\theta_i)$, with $i=1,2$.
 
As for $D=\Rp$, we will admit the existence of self-adjoint extensions
and construct them as maximal symmetric operators. We write them
$H_\varpi(\R)$, where $\varpi$ is a symbolic parameter, the meaning of
which we will clarify further on. We write $\mathcal{B}_\varpi$ the set of
eigenfunctions in the bound spectrum, $\mathcal{F}_\varpi$ that of
eigenfunctions in the free spectrum, $\mathcal{D}_\varpi=\mathcal{B}_\varpi
\bigcup\mathcal{F}_\varpi$ and $\mathcal{S}_\varpi$ the corresponding bound
spectrum.

\subsection{Continuity of probability}
\label{contproba}

Let $\phi_e$ be an eigenfunction of self-adjoint extension $H_\varpi(\R)$. We
will first use the continuity of $\rho(x)=|\phi_e(x)|^2$.

One put apart the case when $\phi_e(0^+)=0$ or $\phi_e(0^-)=0$. Indeed, the
only eigenfunctions which have such limit are the Rydberg ones. In such case,
the continuity of $\rho$ gives $\phi_e(0^+)=\phi_e(0^-)=0$ and $\phi_e$ is
eventually continuous on $\R$.

We recall that \textit{non Rydberg} functions do not cancel at $x=0$. For
such functions, the continuity of $\rho$ implies $|\phi_e(0^-)|=|\phi_e(0^+)|
{\iff}\phi_e(0^-)=\e^{\ii\theta}\phi_e(0^+)$ with $\theta\in[0,2\pi[$.

Let $\phi_{e_1}$ and $\phi_{e_2}\in\mathcal{D}_\varpi$ be two independent
eigenfunctions, $\phi_{e_1}(0^-)=\e^{\ii\theta_1} \phi_{e_1}(0^+)$
and $\phi_{e_2}(0^-)=\e^{\ii\theta_2}\phi_{e_2}(0^+)$.\newline
$|\phi_{e_1}\rangle$ and $|\phi_{e_2}\rangle$ are eigenstates of hermitian
operator $H_\varpi(\R)$, their combination is physical; one can consider state
\newline
$|\psi\rangle=\alpha|\phi_{e_1}\rangle+\beta\e^{\ii\zeta}|\phi_{e_2}\rangle$
with arbitrary coefficients $(\alpha,\beta)\in\R^2$ and $\xi\in[0,2\pi[$. The
evolution in time of $|\psi\rangle$ is given by
\[
|\psi(t)\rangle=
\alpha\e^{-\ii\displaystyle{e_1\hbar t\over2m}}|\phi_{e_1}\rangle+
\beta\e^{\ii\zeta}\e^{-\ii\displaystyle{e_2\hbar t\over2m}}|\phi_{e_2}\rangle\ .
\]
$\rho(x,t)=|\psi(x,t)|^2$ represents a density of probability and must be
continuous with respect to $x$ at all times. One finds
\begin{eqnarray*}
\rho(x,t)
&=&|\alpha|^2|\phi_{e_1}(x)|^2+|\beta|^2|\phi_{e_2}(x)|^2
+2\alpha\beta\;\Re\Big[\overline{\phi_{e_1}(x)}\phi_{e_2}(x)\Big]
\cos\left[{(e_1-e_2)\hbar t\over2m}+\zeta\right]\\
&&-2\alpha\beta\;\Im\Big[\overline{\phi_{e_1}(x)}\phi_{e_2}(x)\Big]
\sin\left[{(e_1-e_2)\hbar t\over2m}+\zeta\right]\ .
\end{eqnarray*}
The continuity of $x\mapsto\rho(x,t)$, valid for all $\alpha$, $\beta$,
$\zeta$ and $t$, implies that of $\Re\big(\overline{\phi_{e_1}}\phi_{e_2}\big)$
and $\Im\big(\overline{\phi_{e_1}}\phi_{e_2}\big)$; so one gets
\[
\overline{\phi_{e_1}(0^+)}\phi_{e_2}(0^+)=
\overline{\phi_{e_1}(0^-)}\phi_{e_2}(0^-)\ .
\]
If one of $\{\phi_{e_1},\phi_{e_2}\}$ is \textit{Rydberg} and cancels at $x=0$,
this relation is always true. If they are both non Rydberg, it reads
$\e^{\ii(\theta_1-\theta_2)}=1\iff \theta_1=\theta_2(2\pi)$, where $(2\pi)$
means modulo $2\pi$.

Eventually,  we have proved the existence of $\theta_\varpi\in[0,2\pi[$ such
that, for all non Rydberg eigenfunctions,
\begin{equation}
\label{epsilon}
\phi_e(0^-)=\e^{\ii\theta_\varpi}\phi_e(0^+)\ .
\ \blacksquare
\end{equation}

\subsection{$\theta$-symmetry}

We still consider $H_\varpi(\R)$. We still assume there exists a non Rydberg
eigenfunction $\phi_e$ in the bound spectrum ($e<0$). From (\ref{fg}), one can
write $\phi_e^>=\mu_k^+f_\eta$ and $\phi_e^<=\mu_k^-\widecheck{f_\eta}$, where
the transposition is defined by $\widecheck{\varphi}(x)=\varphi(-x)$. Then,
(\ref{epsilon}) implies $\mu_k^-=\e^{\ii\theta_\varpi}\mu_k^+$. Thus,
$\phi_e$ is said to be $\theta_\varpi$-symmetrical, where $\theta$-symmetry
is also written $\mathcal{R}(\theta)$ and defined by
\[
\mathcal{R}(\theta)\ :\qquad
\phi_e=\phi_e^>+\e^{\ii\theta}\widecheck{\phi_e^>}\ .
\]
We assume now that there are two or more non Rydberg eigenfunctions in the bound
spectrum, let us write them $\varphi_{k_1}$ and$\varphi_{k_2}$. Note that
$\varphi_{k_1}\;\propto\;\varphi_{k2}\iff
\varphi_{k_1}^>\;\propto\;\varphi_{k2}^>$ (where $\varphi_k^>$ is the
restriction on $\Rp$). Their scalar product reads
\[
\langle\varphi_{k_1}|\varphi_{k_2}\rangle=
2\langle\varphi_{k_1}^>|\varphi_{k_2}^>\rangle\ .
\]
When they are not proportional, $\varphi_{k_1}$ and $\varphi_{k_2}$ can be
eigenfunctions of the same $H_\varpi(\R)$ only if $\varphi_{k_1}^>$ and
$\varphi_{k_2}^>$, their restriction on $\Rp$, are orthogonal each other. From
part~\ref{Rplus}, we get $\omega(\eta_1)=\omega(\eta_2)$. Let us call
$\omega_\varpi$ this constant.  Altogether, we have established the existence
of parameters $\omega_\varpi$ and $\theta_\varpi$, such that all non
Rydberg eigenfunctions $\phi_e$, in the bound spectrum, obey
$\mathcal{R}(\theta_\varpi)$ and $\gb(\eta)=-\omega_\varpi$, with
$\eta=\lambda/(2\sqrt{-e})$, so $\phi_e^>=\varphi_k^>\in
\mathcal{B}_{\omega_\varpi}$. 
$\blacksquare$

We will examine now the situation, where there is also a Rydberg eigenstate in
the domain of $H_\varpi(\R)$, and prove that this Rydberg states has the
opposite symmetry to the non Rydberg one, in the following sense.  Consider
$\phi_{e_1}\in\mathcal{B}_\varpi$, with $\phi_{e_1}(0)\ne0$, and
$\phi_{e_2}\in \mathcal{D}_\varpi$, with $\phi_{e_2}(0)=0$. $\phi_{e_1}$
obeys $\mathcal{R}(\theta_\varpi)$, which reads $\phi_{e_1}^<=
\e^{\ii\theta_\varpi}\phi_{e_1}^>$. One can expand $\phi_{e_2}$ into a
$\theta_\varpi$-symmetrical and a $\theta_\varpi{+}\pi$-symmetrical parts,
$\phi_{e_2}=\phi^{\theta_\varpi}_{e2}+ \phi^{\theta_\varpi+\pi}_{e2}$, as
demonstrated in appendix. Then, one finds
$\phi^{\theta_\varpi}_{e2}=0$, writing
\[
0=\langle\phi_{e_1}|\phi_{e_2}\rangle=
\langle\phi_{e_1}|\phi^{\theta_\varpi}_{e2}\rangle+
\underbrace{\langle\phi_{e_1}|\phi^{\theta_\varpi+\pi}_{e2}\rangle}_{=0}=
2\langle\phi_{e_1}^>|\phi^{\theta_\varpi>}_{e2}\rangle
\]
(the second term is zero by symmetry, cf. appendix) so
$\phi^{\theta_\varpi>}_{e2}$ is orthogonal to $\phi_{e_1}^>$, which is
impossible, since $\phi_{e_1}^>\in \mathcal{D}_{\omega_\varpi}$ and
$\phi^{\theta_\varpi>}_{e2}\in\mathcal{D}_\infty$ because it is a Rydberg
eigenfunction, unless $\phi^{\theta_\varpi>}_{e2} =0$. This proves that
$\phi^{\theta_\varpi}_{e_2}=0$ so $\phi_{e_2}$ obeys
$\mathcal{R}(\theta_\varpi+\pi)$.
$\blacksquare$

Let us examine now free states. We consider a non Rydberg eigenfunction $\phi_e$
with $e>0$. We will find that $\phi_e$ obeys $\mathcal{R}(\theta_\varpi)$ and
that $\phi_e^>=\Psi_k^>\in\mathcal{F}_{\omega_\varpi}$, but the demonstration is
more involved and relies also on the current continuity. To begin with,
following (\ref{FG}), one can write $\phi_e^>=\alpha_k^+F_\eta+\beta_k^+G_\eta$
and $\phi_e^<=\alpha_k^-\widecheck{F_\eta}+\beta_k^-\widecheck{G_\eta}$.
Applying (\ref{epsilon}), one gets
$\beta_k^-=\e^{\ii\theta_\varpi}\beta_k^+$.

\subsection{Conservation of current}

We still consider $H_\varpi(\R)$ and two independent eigenfunctions
$\phi_{e_1}$ and $\phi_{e_2}$ in the domain of $H_\varpi(\R)$ and calculate
the current associated to the mixed state $|\psi(t)\rangle$ defined in section
\ref{contproba}. It becomes, after some calculation,
\begin{eqnarray*}
{\bf j}&=&{\bf j}_1+{\bf j}_2
+{\hbar\alpha\beta\over m}\Re\Bigg[
\overline{\phi_{e_1}(x)}{\partial\phi_{e_2}\over\partial x}(x)-
\phi_{e_2}(x)\overline{{\partial\phi_{e_1}\over\partial x}(x)}\Bigg]
\times\sin\left[{(e_1-e_2)\hbar t\over2m}+\zeta\right]\\
&&\qquad+{\hbar\alpha\beta\over m}\Im\Bigg[
\overline{\phi_{e_1}(x)}{\partial\phi_{e_2}\over\partial x}(x)-
\phi_{e_2}(x)\overline{{\partial\phi_{e_1}\over\partial x}(x)}\Bigg]
\times\cos\left[{(e_1-e_2)\hbar t\over2m}+\zeta\right],
\end{eqnarray*}
where ${\bf j}_1$ and ${\bf j}_2$ are constant. The conservation of probability
law
\begin{equation}
\label{conservj}
{\partial{\bf j}\over\partial x}+{\partial\rho\over\partial t}=0
\end{equation}
applies independently on the sinus and cosine terms, so it eventually reads
\[
\overline{\phi_{e_1}(x)}{\partial^2\phi_{e_2}\over\partial x^2}(x)-
\phi_{e_2}(x)\overline{{\partial^2\phi_{e_1}\over\partial x^2}(x)}
+(e_2-e_1)\,\overline{\phi_{e_1}(x)}\phi_{e_2}(x)=0
\]
and must be verified $\forall x\in\R$. For $x\in\Rp$,
(\ref{conservj})$\iff$(\ref{original}); for $x\in\Rm$, 
(\ref{conservj})$\iff$(\ref{negatif}); so, a particular attention must be paid
to the determination of $\partial{\bf j}/\partial x$ when it is evaluated
through $x=0$. One has
\[
{\partial{\bf j}\over\partial x}(0)=
\lim_{\epsilon_1\to0^+\atop\epsilon_2\to0^+}
{{\bf j}(\epsilon_2)-{\bf j}(-\epsilon_1)\over\epsilon_2+\epsilon_1}\ .
\]
Let us continue the proof concerning non Rydberg free states, which was
sketched in the previous section. We choose $|\phi_{e_1}\rangle$ a non Rydberg
bound state and $|\phi_{e_2}\rangle$ a non Rydberg free one (we assume their
existence; one observes that they are independent). So $\phi_{e_1}^>=
\mu_{k_1}^+f_{\eta_1}$, $\phi_{e_1}^<=\mu_{k_1}^-\widecheck{f_{\eta_1}}$, 
$\phi_{e_2}^>=\alpha_{k_2}^+F_{\eta_2}+\beta_{k_2}^+G_{\eta_2}$ and 
$\phi_{e_2}^<=\alpha_{k_2}^-\widecheck{F_{\eta_2}}
+\beta_{k_2}^-\widecheck{G_{\eta_2}}$, with $\mu_{k_1}^-=
\e^{\ii\theta_\varpi}\mu_{k_1}^+$ and $\beta_{k_2}^-=\e^{\ii\theta_\varpi}
\beta_{k_2}^+$. All terms in the previous limit read ${\hbar\alpha\beta\over m}
\Re(\overline{\alpha}\beta..)\sin(..)+\big(\Re\leftrightarrow\Im\;\&\;
\sin\leftrightarrow\cos\big)$. One applies again the
independence of sinus and cosine, and skips factor $\hbar\alpha\beta\over m$.
The first order of the remaining term reads
\[
{D_{\eta_1}C_{\eta_2}\over2\eta_2\Gamma(1+\eta_1)}
\lim_{\epsilon_1\to0^+\atop\epsilon_2\to0^+}
{\overline{\mu_{k_1}^+}\alpha_{k_2}^+-\overline{\mu_{k_1}^-}\alpha_{k_2}^-\over
\epsilon_1+\epsilon_2}
\]
and exists if and only if $\overline{\mu_{k_1}^+}\alpha_{k_2}^+=
\overline{\mu_{k_1}^-}\alpha_{k_2}^-$ which therefore gives $\alpha_{k_2}^-=
\e^{\ii\theta_\varpi}\alpha_{k_2}^+$.
The second order reads
\[
-{D_{\eta_1}(\eta_1^{\ 2}+\eta_2^{\ 2})\over
4\eta_1^{\ 2}\eta_2^{\ 2}C_{\eta_2}\Gamma(1+\eta_1)}
\lim_{\epsilon_1\to0^+\atop\epsilon_2\to0^+}
{\overline{\mu_{k_1}^+}\beta_{k_2}^+\epsilon_2
+\overline{\mu_{k_1}^-}\beta_{k_2}^-\epsilon_1
\over\epsilon_1+\epsilon_2}
\]
and exists if and only if $\overline{\mu_{k_1}^+}\beta_{k_2}^+=
\overline{\mu_{k_1}^-}\beta_{k_2}^-$ which therefore gives $\beta_{k_2}^-=
\e^{\ii\theta_\varpi}\beta_{k_2}^+$. We have proved that all non Rydberg
obey $\mathcal{R}(\theta_\varpi)$, although we have not determined the set to
which belongs $\phi_e^>$ when $e>0$.
$\blacksquare$
\newpage
Before taking advantage of this result, let us conclude on the current of
probability. For $\phi_{e_1}$ and $\phi_{e_2}$ non Rydberg, \textbf{j} is odd
and the limit of ${\bf j}(x)/x$ when $x\to0$ becomes
\begin{eqnarray*}
{d{\bf j}\over dx}(0)&=&{\overline{\beta_{k_1}^+}\beta_{k_2}^+\over
C_{\eta_1}C_{\eta_2}}\left({1\over(2\eta_2)^2}-{1\over(2\eta_1)^2}\right)
\qquad\hbox{if }e_1>0\hbox{ and }e_2>0\;;\\
&=&{\overline{\mu_{k_1}^+}\beta_{k_2}^+D_{\eta_2}\over
C_{\eta_1}\Gamma(1+\eta_2)}
\left({1\over(2\eta_2)^2}+{1\over(2\eta_1)^2}\right)
\qquad\hbox{if }e_1>0\hbox{ and }e_2<0\;;\\
&=&
{\overline{\mu_{k_1}^+}\mu_{k_2}^+D_{\eta_1}D_{\eta_2}\over
\Gamma(1+\eta_1)\Gamma(1+\eta_2)}
\left({1\over(2\eta_2)^2}-{1\over(2\eta_1)^2}\right)
\qquad\hbox{if }e_1<0\hbox{ and }e_2<0\;.
\end{eqnarray*}
This calculation is valid in both attractive or repulsive cases. For Rydberg
states, the same three limits give  zero (the case $e_1<0$ and $e_2<0$ extends
exactly; the case $e_1>0$ and $e_2<0$ also extends, because the wrong
normalisation vanishes in the zero limit; the case $e_1>0$ and $e_2>0$ is
apart). Altogether, (\ref{conservj}) is respected at all cases.
$\blacksquare$

\subsection{Self-adjoint extensions}

We still consider self-adjoint extension $H_\varpi(\R)$. We assume first that
there exists a non Rydberg bound eigenfunction $\phi_{e_1}$. We have shown that
there are two parameters $\omega_\varpi$ and $\theta_\varpi$ such that it reads
$\phi_{e_1}^>=\varphi_{k_1}^>$ and $\phi_{e_1}^<
=\e^{\ii\theta_\varpi}\widecheck{\varphi_{k_1}^>}$ with
$\gb({\lambda\over2k_1})=-\omega_\varpi$ and $\varphi_{k_1}^>\in
\mathcal{B}_{\omega_\varpi}$. In other words, $\phi_{e_1}$ is a
$\theta$-symmetrical eigenfunction of $\mathcal{B}_\varpi$.

Let us achieve the proof concerning non Rydberg free states; so we assume there
is such an eigenfunction $\phi_{e_2}$, with $e_2>0$. We know $\phi_{e_2}$
obeys $\mathcal{R}(\theta_\varpi)$. So the scalar product
$\langle\phi_{e_1}| \phi_{e_2}\rangle$ reads
\[
0=\langle\varphi_{e_1}|\phi_{e_2}\rangle=
2\langle\phi_{e_1}^>|\phi_{e2}^>\rangle\ ;
\]
it is zero because they are both eigenfunctions of the same operator
$H_\varpi(\R)$. Now, the equality $\widetilde{\mathcal{F}}_{\omega_\varpi}
=\mathcal{F}_{\omega_\varpi}$ implies\newline $\Big\{|\phi_{e_1}^>\rangle\perp
|\phi_{e_2}^>\rangle\iff\phi_{e_2}^>\in\mathcal{F}_{\omega_\varpi}\Big\}$.
This proves that $\phi_{e_2}$ obeys $\mathcal{R}(\theta_\varpi)$ with
$\phi_{e_2}^>\in \mathcal{F}_{\omega_\varpi}$. Altogether, all non Rydberg
states obey $\mathcal{R}(\theta_\varpi)$ with
$\phi_e^>\in\mathcal{D}_{\omega_\varpi}$.
$\blacksquare$

Let us eventually consider any Rydberg eigenfunction $\phi_{e_3}$ of the same
operator $H_\varpi(\R)$. We know that this function is
$\theta_\varpi{+}\pi$-symmetrical.  Reversely, all
$\theta_\varpi{+}\pi$-symmetrical Rydberg eigenfunctions are orthogonal to
any (here non Rydberg) $\theta_\varpi$-symmetrical function (cf. appendix),
so $H_\varpi(\R)$ can be extended into a symmetric operator (one can choose a
trivial action $H_\varpi(\R)|\phi_e\rangle= |0\rangle$), acting on all
eigenstates $\phi_e$ obeying $\mathcal{R}(\theta_\varpi)$ with
$\phi_e^>\in\mathcal{D}_{\omega_\varpi}$ and on all eigenstates $\phi_e$
obeying $\mathcal{R}(\theta_\varpi+\pi)$ with $\phi_e^>\in\mathcal{D}_\infty$.
This extension is maximal by construction and reads
\[
\pi_{\theta_\varpi}\Big(H_{\omega_\varpi}(\Rp)\times 
H_{\omega_\varpi}(\Rm)\Big)\oplus
\pi_{\theta_\varpi+\pi}\Big(H_\infty(\Rp)\times H_\infty(\Rm)\Big),
\]
where $\pi_\theta$ is the projector on $\theta$-symmetrical functions. Since
$H_\varpi(\R)$ is maximal by definition, it is equal to this extension, and
our classification is complete, within the assumption that there are non Rydberg
eigenfunctions (one at least in the bound spectrum, one at least in the free
spectrum). In such case, we define $\varpi=(\omega_\varpi,
\theta_\varpi)$ and our results prove that $\mathcal{D}_\varpi=
\pi_{\theta_\varpi}\big(\mathcal{D}_{\omega_\varpi}\bigcup
\widecheck{\mathcal{D}}_{\omega_\varpi}\big)$
$\oplus\pi_{\theta_\varpi+\pi}\big(\mathcal{D}_\infty \bigcup
\widecheck{\mathcal{D}}_\infty\big)$ and $\mathcal{S}_\varpi= 
\mathcal{S}_{\omega_\varpi}\bigcup
\mathcal{S}_\infty$.
$\blacksquare$

Let us assume now that there are no non Rydberg states. In that case, all
combinations of eigenstates are orthogonal, so the self-adjoint extension is
defined on $\mathcal{D}_\infty\bigcup\widecheck{\mathcal{D}}_\infty$ with no
constraint.  It is maximal by construction, so $H_\varpi(\R)$ equals
$H_\infty(\Rp)\times H_\infty(\Rm)$. We define $\varpi=\infty$ in that
situation. Note that, however, $H_\infty(\R)$ can be identified with
$H_{(\omega,\theta)}(\R)$ for any $\theta\in[0,2\pi[$ and $\omega\to\infty$,
because one can expand any eigenfunction as the sum of its $\theta$-symmetrical
and $\theta{+}\pi$-symmetrical parts.

\subsection{Existence of a non Rydberg bound state}

We consider a self-adjoint extension $H_\varpi(\R)$. We assume there is at
least a non Rydberg eigenstate, otherwise $\varpi=\infty$, which situation
exists and has been studied above.

We can rapidly exclude the situation, where there are no non Rydberg free
eigenstates. Indeed, one knows that all non Rydberg bound states' energies
belong to some set $\mathcal{S}_{\omega_\varpi}$ and that their
eigenfunctions obey $\mathcal{R}_(\theta_\varpi)$, with
$\theta_\varpi\in[0,2\pi[$; so they belong to
$\mathcal{B}_\varpi=\pi_{\theta_\varpi}(\mathcal{B}_\omega\bigcup
\widecheck{\mathcal{B}}_\omega)$. Thus, $H_\varpi(\R)$ can be extended by
$H_{(\omega_\varpi,\theta_\varpi)}(\R)$. Therefore, $H_\varpi(\R)=
H_{(\omega_\varpi,\theta_\varpi)}(\R)$ and there are indeed non Rydberg free
eigenstates.

On the contrary, the situation with non Rydberg free eigenstates and no bound
ones can not be discarded so easily. The demonstration is close to that of
section~\ref{existence}.

We first study the case of a unique non Rydberg free eigenstate
$|\phi_{e_1}\rangle$. There must be a Rydberg free one $|\phi_{e_2}\rangle$ with
$e_2\ne e_1$, otherwise, $H_\varpi(\R)$ would not be physical. (\ref{FG})
reads $\phi_{e_1}^>=\alpha_{k_1}^+F_{\eta_1}+ \beta_{k_1}^+G_{\eta_1}$,
$\phi_{e_1}^<=\alpha_{k_1}^-\widecheck{F_{\eta_1}}+
\beta_{k_1}^-\widecheck{G_{\eta_1}}$,  $\phi_{e_2}^>=\alpha_{k_2}^+F_{\eta_1}$
and $\phi_{e_2}^<=\alpha_{k_2}^-\widecheck{F_{\eta_2}}$.

We have found that there exists $\theta_\varpi$ such that
$\beta_{k_1}^-=\e^{\ii\theta_\varpi} \beta_{k_1}^+$. Then,
$\langle\phi_{e_1}|\phi_{e_2}\rangle=0$ gives
\[
\overline{\beta_{k_1}^+}\alpha_{k_2}^++\overline{\beta_{k_1}^-}\alpha_{k_2}^-=0
\]
so $\alpha_{k_2}^-=-\e^{\ii\theta_\varpi}\alpha_{k_2}^+=
\e^{\ii(\theta_\varpi+\pi)}\alpha_{k_2}^+$. Since the eigenspace associated
to $e_2$ is of dimension 1 (because of the Dirichlet condition, since
$\phi_{e_2}$ is Rydberg), one deduces that $\phi_{e_2}$ obeys
$\mathcal{R}(\theta_\varpi+\pi)$. Thus, from the relation above, $\phi_{e_1}$
obeys $\mathcal{R}(\theta_\varpi)$. Therefore, $\alpha_{k_1}^+/\beta_{k_1}^+=
\alpha_{k_1}^-/\beta_{k_1}^-\equiv\zeta_{k_1}$, one defines $\omega=
-\gf(\eta_1)+{\zeta_{k_1}C_{\eta_1}^{\ \ 2}\over2\eta_1}$, then
$H_\varpi(\R)$ can be extended by $H_{(\omega,\theta_\varpi)}(\R)$, which
admits non Rydberg bound eigenstates. This is indeed contradictory and the case
can be discarded.
$\blacksquare$

Let us assume now there are two independent non Rydberg free states
$|\phi_{e_1}\rangle$ and $|\phi_{e_2}\rangle$.\newline (\ref{FG}) reads
$\phi_{e_1}^>=\alpha_{k_1}^+F_{\eta_1}+ \beta_{k_1}^+G_{\eta_1}$,
$\phi_{e_1}^<=\alpha_{k_1}^-\widecheck{F_{\eta_1}}+
\beta_{k_1}^-\widecheck{G_{\eta_1}}$,
$\phi_{e_2}^>=\alpha_{k_2}^+F_{\eta_1}+\beta_{k_2}^+G_{\eta_2}$
and $\phi_{e_2}^<=\alpha_{k_2}^-\widecheck{F_{\eta_2}}+
\beta_{k_2}^-\widecheck{G_{\eta_2}}$. We have already found that there exist
$\theta_\varpi$ such that $\beta_{k_1}^-=
\e^{\ii\theta_\varpi}\beta_{k_1}^+$ and $\beta_{k_2}^-=
\e^{\ii\theta_\varpi}\beta_{k_2}^+$. We use the continuity of $\bf j$ the
same way as before, constructing a state $|\phi\rangle=\alpha|\phi_{e_1}\rangle+
\beta|\phi_{e_2}\rangle$ and calculating
$\lim_{\epsilon_1\to0\;;\;\epsilon_2\to0}$. One applies again the independence
of sinus and cosine, and skips factor $\hbar\alpha\beta\over m$. Then, the
first order of the remaining term reads
\[
\left(
{C_{\eta_2}\over \eta_1C_{\eta_1}}\big(\overline{\beta_{k_1}^+}\alpha_{k_2}^+
-\overline{\beta_{k_1}^-}\alpha_{k_2}^-\big)+
{C_{\eta_1}\over \eta_2C_{\eta_2}}\big(\overline{\beta_{k_2}^+}\alpha_{k_1}^+
-\overline{\beta_{k_2}^-}\alpha_{k_1}^-\big)\right)
\times{1\over\epsilon_++\epsilon_2}\ ,
\]
so the existence of the limit $\epsilon_1\to0$ and $\epsilon_2\to0$ gives 
\begin{equation}
\label{zetapm}
{\overline{\zeta_{k_1}^+}C_{\eta_1}^{\ \ 2}\over\eta_1}
-{\overline{\zeta_{k_1}^-}C_{\eta_1}^{\ \ 2}\over\eta_1}
={\zeta_{k_2}^+C_{\eta_2}^{\ \ 2}\over\eta_2}
-{\zeta_{k_2}^-C_{\eta_2}^{\ \ 2}\over\eta_2}\ .
\end{equation}
The second order gives $\overline{\beta_{k_1}^+}\beta_{k_2}^+=
\overline{\beta_{k_1}^-}\beta_{k_2}^-$, which one already knows. The scalar
product $\langle\phi_{e_1}| \phi_{e_2}\rangle$ reads
\[
\langle\phi_{e_1}| \phi_{e_2}\rangle=
{2\lambda\over k_1^{\ 2}-k_2^{\ 2}}\Bigg(
{C_{\eta_2}\over C_{\eta_1}}{\overline{\beta_{k_1}^+}\alpha_{k_2}^+
-\overline{\beta_{k_1}^-}\alpha_{k_2}^-\over2\,\eta_1}-
{C_{\eta_1}\over C_{\eta_2}}
\times{\overline{\beta_{k_2}^+}\alpha_{k_1}^+
-\overline{\beta_{k_2}^-}\alpha_{k_1}^-\over2\,\eta_2}+
{\gf(\eta_1)-\gf(\eta_2)\over C_{\eta_1}C_{\eta_2}}
\big(\overline{\beta_{k_1}^+}\beta_{k_2}^+
+\overline{\beta_{k_1}^-}\beta_{k_2}^-\big)\Bigg)
\]
thus $\langle\phi_{e_1}| \phi_{e_2}\rangle=0$ gives 
\begin{equation}
\label{zetamp}
{\overline{\zeta_{k_1}^+}C_{\eta_1}^{\ \ 2}\over\eta_1}
+{\overline{\zeta_{k_1}^-}C_{\eta_1}^{\ \ 2}\over\eta_1}-2\gf(\eta_1)
={\zeta_{k_2}^+C_{\eta_2}^{\ \ 2}\over\eta_2}
+{\zeta_{k_2}^-C_{\eta_2}^{\ \ 2}\over\eta_2}-2\gf(\eta_2)\ .
\end{equation}
Adding (\ref{zetapm}) and (\ref{zetamp}) proves that $(\zeta_{k_1}^+,
\zeta_{k_2}^+)$ obeys (\ref{boundstate}); thus, $\zeta_{k_i}^+$ are real and
obey (\ref{zeta}). Subtracting (\ref{zetapm}) and (\ref{zetamp}) proves that
$(\zeta_{k_1}^-,\zeta_{k_2}^-)$ obeys (\ref{boundstate}); thus, $\zeta_{k_i}^-$
are real and obey (\ref{zeta}). Then, (\ref{zetamp}) proves that the same
$\omega$ can be associated to all functions $\phi_{k_1}^>$, $\phi_{k_1}^<$,
$\phi_{k_2}^>$ and $\phi_{k_2}^<$.  Therefore, notwithstanding we did not
establish $\zeta_i^+=\zeta_i^-$, one can introduce any bound state associated to
$\varphi_\eta$ with $\gb(\eta)=-\omega$, and extend the action of $H_\varpi(\R)$
on these bound states, keeping the operator symmetric. This is contradictory, so
the result is proved.
$\blacksquare$

\subsection{Discussion}

\subsubsection*{Mathematical interpretation}

We have determined all self-adjoint extensions. $\theta$-symmetrical states obey
$\mathcal{C}(\theta)$. So, all eigenfunctions $\phi_e$ of
$H_{\omega,\theta}(\R)$, respecting $\mathcal{R}(\theta)$, with $\phi_e^>\in
\mathcal{D}_\omega$, obey $\mathcal{C}(\theta)$ and all eigenfunctions $\phi_e$
of $H_{\omega,\theta}(\R)$, respecting $\mathcal{R}(\theta+\pi)$, with
$\phi_e^>\in \mathcal{D}_\infty$, obey $\mathcal{C}(\theta+\pi)$. For
$H_\infty(\R)$, $\mathcal{C}(\theta)$ reduces to Dirichlet conditions (all
eigenfunctions obey $\mathcal{C}(\theta')$ for any $\theta'\in[0,2\pi[$). 

\subsubsection*{The Dirichlet case in one dimension}

We focus on the case $\varpi=\infty$ and study eigenfunctions $\phi_e$.  Both
attractive and repulsive case can be considered, but we will focus on the first
one.

Let us consider the bound spectrum. From what precedes, $\mu$ in
(\ref{description}) is entirely free. Therefore,
$\{|\phi_e^>\rangle,|\phi_e^<\rangle\}$  is a basis of the eigenspace $E_e$
corresponding to energy $e$.  This is an exceptional violation of the general
result, which asserts that an energy in the bound spectrum is non degenerated in
one dimensional systems. Here, the eigenspace $E_e$ has dimension 2.  However,
examining the standard demonstration\cite{Aslangul}, on observes that it is
based on a Wronskian theorem, which can not apply here.

Another basis is composed of $\{|\phi_e^+\rangle,|\phi_e^\textrm{-}\rangle\}$,
the even and odd extensions on $\R$.  For $\omega=\infty$, one observes that
$\phi_e^-\in C(\R)$ for all $e\in\mathcal{S}_\infty\bigcup\R_+$.
$H_\infty(\R)$, defined on these basis, is closed and therefore self-adjoint.
More generally, one can use
$\{|\phi_e^\theta\rangle,|\phi_e^{\theta+\pi}\rangle\}$, for any
$\theta\in[0,2\pi[$.

\section{Physical applications}

We study different possible extensions of this work to real physical situations.

\subsection{The hydrogenate case in three dimension}

Let us focus on the case $\D=\R^3$, using the mapping $\Phi(r)=\phi(r)/r$, where
$\phi$ is the one-dimensional solution and $\Phi$ the radial part of the
three-dimensional wavefunction. We will only consider the attractive case here.

Let us connect our parametrization $\omega$ with that of$\!$
{\setcitestyle{numbers,open={},close={},comma}
Ref.~\cite{Albeverio}}, which parameter is written
$\alpha$. We will show the connection for bound states only, but this can be
done for all states. The first order expansion of any state $\phi_e$ with $e<0$
reads
\[
\phi_e(x)=a+\lambda a x\ln(|\lambda|x)+b x\ ;
\]
this expression holds both in attractive and repulsive cases. $\omega$ can be
expressed in terms of $b/a$, which reads
\[
\omega={1\over|\lambda|}({b\over a}-\lambda)\ .
\]
In$\!$ {\setcitestyle{numbers,open={},close={},comma}
Ref.~\cite{Albeverio}}, where $\lambda$ reads
$\gamma$, one finds parameters $\phi_0=a$ and $\phi_1=b$, so one gets
\[
\alpha={1\over4\pi}{b\over a}\quad\iff\quad
\omega={1\over|\lambda|}(4\pi\alpha-\lambda)\ .
\]
As it is well known\cite{Cohen}, for $L>0$, the solutions of the Schrödinger
equation which do not cancel at $r=0$ do not belong to $L^2(\R^3)$ and must
therefore be discarded. On the contrary, that, corresponding to the case $L=0$,
belong to $L^2(\R^3)$ (all $g_\eta$ solutions, which diverge at $r\to\infty$ are
excluded from this discussion). This is the reason why the $L\ne0$ subspaces
appearing in (2.1.13) of$\!$
{\setcitestyle{numbers,open={},close={},comma} Ref.~\cite{Albeverio}} have no
parametrization, contrary to the $L=0$ one.

This helps us interpreting what these authors mean by \og$H_{\gamma,\alpha,y}$
describes the Coulomb interaction plus an additional point interaction\fg: the
eigenfunctions for $\alpha<\infty$ are divergent eigenfunctions and not
physical, although they belong to $L^2(\R^3)$, so they do not describe the
physical Coulomb interaction. Most authors have similarly assumed that the only
admissible Coulomb bound states are the Rydberg ones, given by the Laguerre
polynomial
\[
\Phi(r)=\sqrt{2\lambda\over n^{3/2}}\e^{-r} L'_n(2r)
\]
with a specific normalization (assuming that the spherical function reads
$1/\sqrt{4\pi}$ for kinetic momentum $L=0$). This solution exactly corresponds
to the $\omega=\infty$ Dirichlet case, which is also the $\alpha=\infty$ one.

Actually, no fundamental principle of quantum mechanics justifies discarding
solutions that diverge for $r\to0$, since the probability
$\int|\Phi(r)|^2r^2dr$ is finite (in the basic meaning ``not infinite'').
However, experimental evidences, from the original Rydberg spectrum, are in
excellent agreement with this assumption. We find that experimental data%
\cite{Herzberg} are only compatible with $|\omega|>27779$. We have simply
compared the ratio $E_2-E_m\over E_2-E_n$, for several $(m,n)$ couples, as
determined from these data, with that calculated from the exact values of
$\mathcal{S}_\omega$. Actually, $(m,n)=(5,3)$ gives the highest (best) limit of
possible values for $\omega$.

Based on these physical grounds, we will follow the common choice and, dealing
with the case $D=\R^3$, discard all divergent wavefunctions, therefore reducing
the parameter range to $\omega=\infty$, the self-adjoint extension corresponding
to Dirichlet solutions. We can justify this choice, from a mathematical point of
view, by reminding that the deficiency coefficient of $H(\R^3)$ is zero. We will
discuss this point further on.

\subsection{Explicit spectra for a semi-infinite line}

\begin{figure}[t]
\vglue-1.0cm
\begin{center}
\includegraphics[width=8cm]{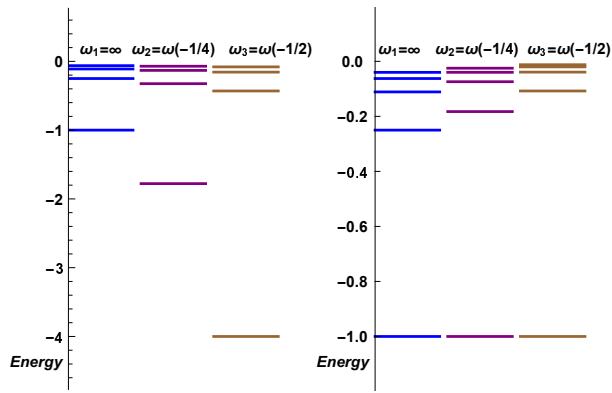}
\end{center}
\caption{Spectra for $\omega_i$ ($\scriptstyle i=1,2,3$) defined in the text. On
the left, we show the absolute values, on the right, we normalize energies so
that the lowest energy is $-1$. The variation of $E_{n+1}-E_n$, when $n$ is
increased, is steeper for $\omega_3$.}
\label{spectresBR}
\end{figure}

The calculated spectra $\mathcal{S}_\omega(\Rp)$ vary significantly, for
different values of $\omega$. We show three of them in Fig.~\ref{spectresBR},
corresponding to $\omega_1=\infty$ (Rydberg spectrum),
$\omega_2=\omega(-1/4)\approx2.3$ and $\omega_3=\omega(-1/2) \approx{-}0.27$
(close to the Neumann case). As already pointed out, in any one-dimensional
system the experimental determination of this spectra would allow that of the
limit $\phi'(0^+)/\phi(0^+)$ in the vicinity of the charge defect.
\noindent
Turning back to the one-dimensional attractive case, defined in $\D=\Rp$, one
observes that the previous physical arguments used in the three-dimensional case
cannot apply, because no wavefunction is ever diverging. Therefore, one must
consider all parameters, $\omega\in]-\infty,\infty[$ in the attractive case,
$\omega\in]-\infty,2\geul]$ in the repulsive one.

The determination of $\omega$ is highly system dependent.  If any experimental
spectrum, close enough to this case, can be measured in the future with high
enough precision, like that of a one-dimensional quantum wire (like a carbon
nanotube) with a charge defect at one extremity or an hydrogen atom in very
intense magnetic field, then we argue that the limit condition
$\phi'(0)/\phi(0)$, at that extremity, will be determined by examining the
sequence of energies $E_1<E_2<E_3$... and in particular the sequence of their
ratio.

\subsection{Regularization of the potential}

We consider here the regularized potential $V_\varepsilon=
{\lambda/\sqrt{x^2+\varepsilon^2}})$ in the attractive case, with $\D=\R$. This
is a way to address the $1+\varepsilon$-dimensional case, since this potential
describes the situation where the charge is lightly displaced from axis $\R$ in
the 3-dimensional space. When
$\varepsilon\to0$, it converges towards the Coulomb potential,
$V_\varepsilon\to V$. We focus on the negative (bound) spectrum of the
corresponding Hamiltonian $H_\varepsilon(\R)$, which is self-adjoint.

\begin{figure}[H]
\centering
\includegraphics[width=6cm]{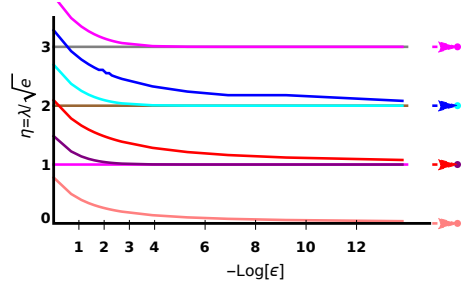}
\caption{First energies of ${p^2\over2m}+V_\varepsilon$
($e_{\frac12}^\varepsilon$, $e_1^\varepsilon$, $e_{\frac32}^\varepsilon$,
$e_2^\varepsilon$, $e_{\frac52}^\varepsilon$ and $e_3^\varepsilon$ from bottom
to top) versus $\ln(1/\varepsilon)$ in dimensionless $y$-scale. The asymptotic
limit is indicated by an arrow on the right, for each curve and by the
horizontal straight lines.}
\label{spectre}
\end{figure}

This spectrum is found discrete and non degenerate $\forall\varepsilon\ne0$.  In
this case, all eigenfunctions are orthogonal and form a complete basis, because
they obey to $H(\R^3)$, which is self-adjoint, as explained before. They
separate into two groups, odd functions $\chi^\varepsilon_{2p}$ with
$p\in\N^\ast$, and even ones $\chi^\varepsilon_{2p+1}$ with $p\in\N$. We will
note $e_p^\varepsilon$ the energies corresponding to odd solutions, and
$e_{p+{1\over2}}^\varepsilon$ that of even solutions. Fig.~\ref{spectre} shows
the first (smallest) energies as a function of $\ln1/\varepsilon$.
\noindent
When $\varepsilon\to0$, even wavefunctions $\chi^\varepsilon_{2p}\to\chi^0_{2p}=
\phi_{-\lambda^2/(4p^2)}$ while their energy rapidly reaches $-\lambda^2/(4p^2)$
the corresponding Rydberg energy. Odd ones also $\chi^\varepsilon_{2p+1}\to
\phi_{-\lambda^2/(4p^2)}$ while their energy reaches $-\lambda^2/(4p^2)$ the
Rydberg energy. This is conform with the 2-degeneracy that is proved in the case
$\varpi=\{\infty^+,\infty^-\}$, which shows that $H_\varepsilon(\R)\to
H_\infty(\R)$. An odd eigenfunction seems to be converging towards a zero energy
eigenfunction, but it vanishes as $\varepsilon\to0$, in conformity with our
discussion about these functions.

\section{Spectral theorem}

We discuss the way one should write the spectral theorem, in the case of
incompatible self-adjoint extensions.

\subsection{Spectral theorem in $\Rp$}

For each value $\omega$, $H_\omega(\Rp)$ is self-adjoint, so the spectral
theorem is valid. Therefore, any function $\psi_i\in L^2(\Rp)$ can be developed
on the basis $\mathcal{B}_\omega(\Rp)\bigcup$ $\mathcal{F}_\omega(\Rp)$
\[
\psi_i(x)=\sum_{\eta\in\mathcal{S}_\omega}b^i_k\varphi_k(k x)
+\int_{\R_+}c^i_k\Psi_k(kx){dk\over\pi}
\hbox{with }b^i_k=\langle\varphi_k|\psi_i\rangle\quad\hbox{and }\
c^i_k=\langle\Psi_k|\psi_i\rangle\ .
\]
end{eqnarray*}
For $\omega=\infty$ and $\lambda<0$ (attractive case), this formula is
equivalent to Eq. 19.171, in$\!$
{\setcitestyle{numbers,open={},close={},comma} Ref.~\cite{Aslangul}} with
a different normalization (we preferred to use $k$ parameter, rather than $E$).
We have checked this formula numerically on several examples,
$x\mapsto\e^{-x^2}$, $x\mapsto x\e^{-x^2}$, etc. One can, in particular, expand
a function $\varphi_k$, with $\omega({\lambda\over2k})\ne\omega_1$ on
$\mathcal{B}_{\omega_1}(\Rp)\bigcup \mathcal{F}_{\omega_1}(\Rp)$, which we have
done for functions $\psi_0=\varphi_{-\lambda}$ (setting
$\omega_0=\omega(-{1\over2})$) or $\psi_0=\varphi_{-{\lambda\over3}}$ (setting
$\omega_0=\omega(-{3\over2})$), while choosing $\omega_1=\infty$.

$\omega=\omega_0$ or $\omega=\omega_1$, is well defined by this expansion,
writing for instance
\[
H_{\omega_1}(\Rp)|\psi_0\rangle=-\sum_{-k_1^{\ 2}\in\mathcal{S}_{\omega_1}}
k_1^{\ 2}b^0_{k_1}\varphi_{k_1}(k_1 x)
+\int_{\R_+}k_1^{\ 2}c^0_{k_1}\Psi_{k_1}(k_1x){dk_1\over\pi}\ .
\]
This result differs from $H_{\omega_0}(\Rp)|\psi_0\rangle$, which reads
\[
H_{\omega_0}(\Rp)|\psi_0\rangle=-k_0^{\ 2}|\psi_0\rangle
=-\sum_{-k_1^{\ 2}\in\mathcal{S}_{\omega_1}}
k_0^{\ 2}b^0_{k_1}\varphi_{k_1}(k_1 x)
-\int_{\R_+}k_0^{\ 2}c^0_{k_1}\Psi_{k_1}(k_1x){dk_1\over\pi}\ .
\]

Finally, one should be aware that, as a formal derivative operator, the action
of $H(\Rp)$ on $\psi_0$ is well defined. In particular, one is interested by its
action on eigenfunctions $\phi_e$. One eventually finds
\[
H(\Rp)|\phi_e\rangle=e|\phi_e\rangle
\]
which means that $H(\Rp)$ acts on $\psi_e$ as $H_\omega(\Rp)$ with
$\omega=\omega(e)$, the index of energy $e$. However, $H(\Rp)$ is not a
\textit{good} operator, because it does not correspond to the same self-adjoint
extension, for each state.

Technically, the last result can be understood as follows: $d/\!dx$ does not
commute with $\int dk$ in the former development. Indeed, when the derivation is
performed \textbf{inside} the integral, it produces a factor $\eta\;\propto\;
1/k$ which makes it improper.

This analysis is common with that, which can be made for $H=-d^2/dx^2$;
the divergence of the Coulomb potential is not entirely responsible of
the loss of  self-adjointness.

\subsection{Spectral theorem in $\R$}

The spectral theorem in $\R$ can be formulated after that in $\Rp$. Each
$\theta$-symmetrical and $\theta{+}\pi$-symmetrical part of any function can be
expanded separately. Considering $H_\varpi(\R)$, with $\varpi=
(\omega,\theta)$, any function $\phi\in L^2(\R)$ expands into $\phi=\phi^\theta
+\phi^{\theta+\pi}$.  Then $\phi^\theta$ expands in $\mathcal{B}_{\omega}(\R)
\bigcup\mathcal{F}_{\omega}(\R)$ exactly as $\phi^{\theta>}$ in
$\mathcal{B}_{\omega}(\Rp)\bigcup\mathcal{F}_{\omega}(\Rp)$ but for a
supplementary factor $1/\sqrt{2}$ : one should take the expansion calculated for
$\D=\Rp$ and allow $x\in\R$; similarly $\phi^{\theta+\pi}$ expands in
$\mathcal{B}_{\infty}(\R) \bigcup\mathcal{F}_{\infty}(\R)$ exactly as
$\phi^{\theta+\pi>}$ in
$\mathcal{B}_{\infty}(\Rp)\bigcup\mathcal{F}_{\infty}(\Rp)$ but for a
supplementary factor $1/\sqrt{2}$.

It applies also in the particular case $\varpi=\infty$, choosing any
arbitrary $\theta$. In this case, one can also write $f=f^>+f^<$ (where $f^>$
extends in $\Rm$ as zero and $f^<$ extends in $\Rp$ as zero). $f^>$ expands in
$\mathcal{B}_\infty(\Rp)\bigcup\mathcal{F}_\infty(\Rp)$ and $f^<$ expands in
$\mathcal{B}_\infty(\Rm)\bigcup\mathcal{F}_\infty(\Rm)$. This is the right place
to observe that $\mu$, defined in (\ref{description}), is not determinate in
this particular case. One can indeed choose $\mu=0$ (i.e. $f=f^>$) or
$\mu=\infty$ (i.e. $f=f^<$). We discuss this supplementary degree of freedom
further.

\section{Topological classification of the extension parameter space}

\subsection{Structure for $\D=\Rp$ in the repulsive case}

The structure of the order parameter seems to be equivalent to the interval
$]-\infty,2\geul]$ in the repulsive case, which is topologically equivalent to
interval $]0,1]$. This is notwithstanding the special case $H_{2\geul}(\Rp)$,
which we found for the zero energy. This case corresponds to $\omega=2\geul$,
but, what should now be pointed out is that the regular limit $\omega\to2\geul$,
which can be constructed, using $g_\eta$, does not exist. One finds indeed that
eigenfunction $g_\eta$ tends to a singular distribution with $\{0\}$ support.
Looking for such a solution, one substitutes again $\varphi_0=\sum_{n=0}^\infty
a_n\delta^{(n)}$ in (\ref{original}). When $\eta\to\infty$, one finds that all
coefficients $a_n$ are zero. 

This indicates that the right boundary of $]{-}\infty,2\geul]$ is apart, one
should write, instead, $]{-}\infty,2\geul[\bigcup\{2\geul\}$ and draw
$\stackrel{\hfill^\bullet}{\rule{0.8cm}{0.5mm}}$ to characterize this space,
which is topologically equivalent to $\Rd$.

\subsection{$U(1)$ structure for $\D=\Rp$ in the attractive case}
\label{U1struct}

\begin{figure}[H]
\begin{center}
\includegraphics[width=8cm]{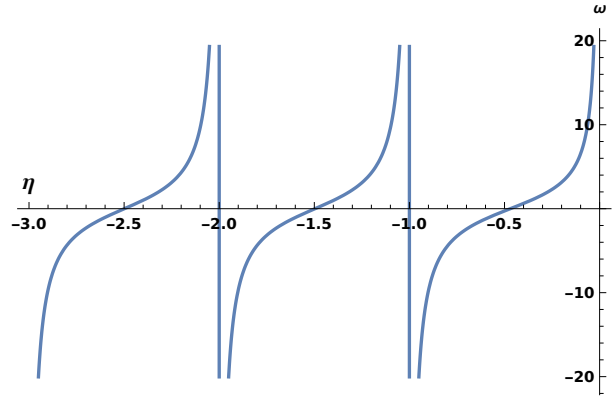}
\end{center}
\caption{$\omega$ versus $\eta$ in the attractive case.}
\label{fOliv}
\end{figure}

Let us observe on Fig.~\ref{fOliv} the curve of index $\omega(\eta)$ in the
attractive case. It is not periodic, but there is an infinity of vertical
asymptotes at positions $\eta=-n$, $n\in\N^\ast$.
\noindent
Any interval $[-n-1,-n]$, for $n\in\N$, covers all indices $\omega$. In other
words, any eigenfunction $\varphi_\eta\in L^2(\Rp)$, with index $\eta$, belongs
to the bound spectrum of $H_\omega(\Rp)$, where $\omega(\eta)$ is determined by
this curve.  $\omega\to-\infty$ and $\omega\to\infty$ are identified (to the
Rydberg solutions), which proves that the set of all extensions of $H(\Rp)$ is
mapped on a space, which is topologically equivalent to the circle $U(1)$.

\subsection{Structure for $\D=\R$}

It is worth pointing out that, although the space of extension parameter is
reduced, as a consequence of the continuity condition at $x=0$, we get the same
deficiency coefficients as Oliveira \textit{et al}\cite{Oliveira} in that
$\D=\R$ case, which are (2,2).

\begin{figure}[t]
\vglue-4.5cm
\begin{center}
\includegraphics[width=10cm]{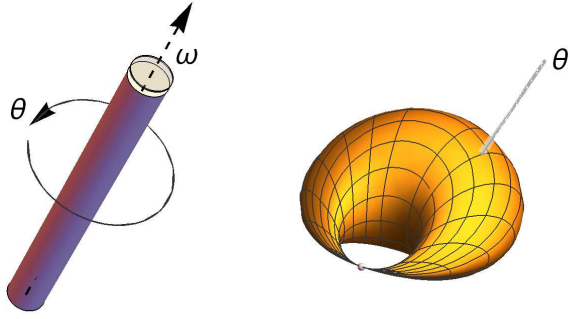}
\end{center}
\vglue-3.9cm
\caption{Representation of the order parameter space in the repulsive case
(left) or attractive case (right). Left are represented the $\omega=2\geul$
closing circle, the $\omega$ axis (which is supposed to vary from $-\infty$ to
$2\geul$) and the gauge parameter $\theta$. Right is represented the
$\omega=\pm\infty$ point at the strangling point and a $\theta$-circle is
pointed out: all  orthogonal lines to this circle vary with $\omega$.}
\label{topol}
\end{figure}

Since there is no Rydberg state in the repulsive case, the structure due to
parameters $(\omega,\theta)$ is very simple, it is an infinite cylinder
$(\R,+\infty)\times[0,2\pi[$, with a closed boundary at one side, as represented
in Fig.~\ref{topol}.
\noindent
The structure in the attractive case is more like a torus, with a strangling,
that is a singular point of infinitely small narrowness, corresponding to
$\omega=\pm\infty$, as seen on Fig.~\ref{topol}.

The $\theta$-symmetry introduces a phase factor $\pm\e^{\ii\theta}$ when a
particle passes $x=0$. Factor  $\e^{\ii\theta}$ is arbitrary but identical for
all states associated to $H_{(\omega,\theta)}(\R)$, similarly to standard gauge
symmetry.

\section{Conclusion}

The one-dimensional Schrödinger equation with a Coulomb $\lambda/|x|$ potential
brings unusual difficulties, for the physical interpretation of its solutions.
Indeed, the corresponding hamiltonians $H(\Rp)$ and $H(\R)$ admit an infinity of
self-adjoint extensions, classified by a real parameter $\omega$. In the case of
$H(\Rp)$ with an attractive Coulomb potential, $\omega$ is defined in the space
$\R$ where $-\infty$ is identified with $\infty$; this space is topologically
equivalent to the circle $U(1)$. In the case of $H(\Rp)$ with a repulsive
Coulomb potential, $\omega$ is defined in $]-\infty,2\geul]$. In both cases,
parameter $\omega$ must be chosen according to the limit
${\partial\phi(x)\over|\lambda|\partial x}/\phi(x) \pm\ln(|\lambda|x)$ when
$x\to0$, where $\pm$ is the sign of $\lambda$. In the attractive case, the
particular value $\omega=\infty$ brings the Dirichlet solutions, which obey
$\phi(0)=0$ and correspond to the standard Rydberg spectrum, while the other
spectra are unusual and have never been observed yet. In the repulsive case, the
particular value $\omega=2\geul$ gives a continuous spectrum $R_+$, the zero
energy eigenfunction of which is bounded.

In the case of $H(\R)$, physical constraints yield a phase gauge $\theta$, which
describes the discontinuity of wavefunctions at $x=0$. If the Coulomb potential
is attractive, two situations may occur: either one finds two separate spectra,
the eigenstates of which are orthogonal and obey, respectively,
$\mathcal{R}(\theta)$ and $\mathcal{R}(\theta+{\pi\over2})$ symmetry; or the
spectrum is the standard Rydberg one, with an exceptional 2-degeneracy of all
eigenfunctions. We did not study the repulsive case here, but we induce that
there is also a supplementary symmetry $\mathcal{R}(\theta)$, giving the
representation sketched in Fig.~\ref{topol} (left).

This study brings up new considerations about quantum physics: in order to
conciliate the classification of $H(\Rp)$ and $H(\R^3)$ with standard
experimental measures of the hydrogen electronic energy levels, one has to
discard all \textbf{divergent} wavefunctions, but we could not justify this
choice. So we suggest to add a postulate in quantum physics, stipulating that no
divergent wavefunction can be admitted, in other words all wavefunctions are
bounded. Indeed, this would give an explanation why one never observes any
physical states with $\omega\ne\infty$.

This work shows that one must be very careful when using the spectral theorem
for an unbounded hamiltonian. At a time when theoretical physics research
includes new and mathematically unexpected objects (like complex eigenvalues for
hamiltonians, skyrmions, Majorana fermions), advanced studies of non
self-adjoint hamiltonians are necessary, and, what seemed old-fashioned physics
reveals an essential source of inspiration and comprehension, to
determinate whether a self-adjoint extension is valid or not.

\appendix
\section{Appendix}

\subsection{Overlappings}

Some overlappings have been calculated in$\!$
{\setcitestyle{numbers,open={},close={},comma}
Ref.~\cite{abramovici}}. We give here a general
derivation of all scalar products. Let us first consider the case of a
bound-unbound product $\langle\varphi_{k_1}|\Psi_{k_2}\rangle$. One has
\begin{eqnarray*}
\int_0^\infty dx\,
\varphi_{k_1}(x){\partial^2\Psi_{k_2}\over\partial x^2}(x){-}
\Psi_{k_2}(x){\partial^2\varphi_{k_1}\over\partial x^2}(x)
&=&
\left[\!\varphi_{k_1}(x){\partial\Psi_{k_2}\over\partial x}(x){-}
\Psi_{k_2}(x){\partial\varphi_{k_1}\over\partial x}(x)\!\right]_{0^+}^\infty\\
\hbox{substituting (\ref{original}) for $\Psi$ or $\varphi$, one gets}
&=&
-(k_1^{\ 2}+k_2^{\ 2})\int_0^\infty dx\;\varphi_{k_1}(x)\Psi_{k_2}(x)
\end{eqnarray*}
which leads to (\ref{scfF}). The demonstration for a bound-bound product
$\langle\varphi_{k_1}|\varphi_{k_2}\rangle$ is very similar. One has
\begin{eqnarray*}
\int_0^\infty dx
\varphi_{k_1}(x){\partial^2\varphi_{k_2}\over\partial x^2}(x){-}
\varphi_{k_2}(x){\partial^2\varphi_{k_1}\over\partial x^2}(x)&=&
\left[\!\varphi_{k_1}(x){\partial\varphi_{k_2}\over\partial x}(x){-}
\varphi_{k_2}(x){\partial\varphi_{k_1}\over\partial x}(x)\!\right]_{0^+}^\infty
\\
\hbox{substituting (\ref{original}), one gets}
&=&(k_2^{\ 2}-k_1^{\ 2})\int_0^\infty dx\; \varphi_{k_1}(x)\varphi_{k_2}(x)
\end{eqnarray*}
which leads to (\ref{scal}).

Using the same method, we can verify the normalization of free states. One has
\begin{eqnarray*}
\int_0^\infty dx
\Psi_{k_1}(x){\partial^2\Psi_{k_2}\over\partial x^2}(x)-
\Psi_{k_2}(x){\partial^2\Psi_{k_1}\over\partial x^2}(x)&=&
\left[\!\Psi_{k_1}(x){\partial\Psi_{k_2}\over\partial x}(x)-
\Psi_{k_2}(x){\partial\Psi_{k_1}\over\partial x}(x)\right]_{0^+}^\infty\\
\hbox{substituting (\ref{original}), one gets}
&=&(k_1^{\ 2}-k_2^{\ 2})\int_0^\infty dx\; \Psi_{k_1}(x)\Psi_{k_2}(x)i\;.
\end{eqnarray*}
While, in previous calculations, the only non zero contribution of the $[\ ]$
lies at $x=0^+$, here, both this boundary and the infinite one contributes. We
have not given in (\ref{scFF}) the detailed calculation of this last
contribution. Let us replace $\infty$ by $L$ and take $L\to\infty$ afterwards.
For $\langle F_{\eta_1}|F_{\eta_2}\rangle$ and $\langle G_{\eta_1}|
G_{\eta_2}\rangle$, one finds a main contribution
\begin{eqnarray*}
m_{\eta_1\eta_2}&\equiv&
\sin\big(k_1L-k_2L-\eta_1\ln(2k_1L)+\eta_2\ln(2k_2L)
+\Gamma(1{+}\ii\eta_1)-\Gamma(1{+}\ii\eta_2)\big)\\
&=&\sin((k_1-k_2)A_{\eta_1\eta_2})\cos(s_{\eta_1\eta_2})+
\cos((k_1-k_2)A_{\eta_1\eta_2})\sin(s_{\eta_1\eta_2})\ ,
\end{eqnarray*}
with $A_{\eta_1\eta_2}{=}L+2{\eta_1\eta_2\over\lambda}\ln(|\lambda|L)$ and
$s_{\eta_1\eta_2}{=}\eta_1\ln|\eta_1|-\eta_2\ln|\eta_2|+\Gamma(1{+}\ii\eta_1)$
$-\Gamma(1{+}\ii\eta_2)$. The other contributions expand as power of $1/L$ and
are suppressed when $L\to\infty$. Similarly,
$\langle G_{\eta_1}|F_{\eta_2}\rangle$ gives a contribution 
\begin{eqnarray*}
n_{\eta_1\eta_2}&\equiv&
\cos\big(k_1L-k_2L-\eta_1\ln(2k_1L)+\eta_2\ln(2k_2L)
+\Gamma(1{+}\ii\eta_1)-\Gamma(1{+}\ii\eta_2)\big)\\
&=&\cos((k_1-k_2)A_{\eta_1\eta_2})\cos(s_{\eta_1\eta_2})-
\sin((k_1-k_2)A_{\eta_1\eta_2})\sin(s_{\eta_1\eta_2})\ ,
\end{eqnarray*}
while $\langle F_{\eta_1}|G_{\eta_2}\rangle$ gives $-n_{\eta_1\eta_2}$.  The
other contribution also cancel as powers of $1/L$. Note that
$A_{\eta_1\eta_2}\to\infty$ when $L\to\infty$, whatever $\eta_1$ and $\eta_2$,
so we will write $A$ for short.

Eventually, taking into account all factors $\alpha_{\eta_1}^\omega$,
$\alpha_{\eta_2}^\omega$, $\beta_{\eta_1}^\omega$ and $\beta_{\eta_2}^\omega$
and $1/(k_1^{\ 2}-k_2^{\ 2})$, one finds exactly three kinds of contribution,
\[
(\alpha_{\eta_1}^\omega\alpha_{\eta_2}^\omega+\beta_{\eta_1}^\omega
\beta_{\eta_2}^\omega)\sin((k_1-k_2)A)/(k_1-k_2)\;,\qquad
z(k_1,k_2)\sin(A(k_1-k_2))\quad\hbox{and}\quad
z(k_1,k_2)\cos(A(k_1-k_2))
\]
where $z$ are non given continuous functions, while the first kind occurs only
once and may be rewritten\newline
$\big((\alpha_{\eta_1}^\omega)^2+(\beta_{\eta_1}^\omega)^2\big)
\sin((k_1-k_2)A)/(k_1-k_2)$, since the difference is of the second kind (note
that all these functions $z$ arise from a cancellation of the $k_1-k_2$ factor
in the denominator with a similar factor in the numerator arising from
Young-Taylor expansion). One must take $|\alpha_{\eta_1}^\omega|^2+
|\beta_{\eta_1}^\omega|^2=1$ to get generalized orthonormality.  Taking into
account the $dk/\pi$ integration factor (which means dividing by $\pi$), on
finds $\sin((k_1-k_2)A)/(k_1-k_2)\to\delta(k_1-k_2)$ when $A\to\infty$. Both
$\sin(A(k_1-k_2))\to0$ and $\cos(A(k_1-k_2))\to0$ as $L\to\infty$, so all the
contribution of the second and third kind are suppressed at this limit. Thus our
calculations are proved.

\subsection{Decomposition of a function}

For all $\theta\in[0,2\pi[$, any function $f$, $\R\to\C$, can be decomposed into
$\theta$-symmetrical and $\theta+\pi$-symmetrical parts,
$f=f_\theta+f_{\theta+\pi}$. The demonstration is similar to that of the
decomposition into even and odd parts. We write $f^>$ the restriction of $f$ on
$\Rp$ and $f^<$ that on $\Rm$. Then, one finds
\begin{eqnarray*}
f_\theta(x)=
\left\{\begin{array}{ll}
f^>_\theta(x)&\forall x>0\;;\\
f^<_\theta(x)&\forall x<0\;;
\end{array}\right.\quad
&\hbox{with}&\quad
f^>_\theta=
{\displaystyle f^>{+}\e^{\ii\theta}\widecheck{f^<}\over\displaystyle2}\ ;
f^<_\theta=
{\displaystyle f^<{+}\e^{-\ii\theta}\widecheck{f^>}\over\displaystyle2}\ ;\\
f_{\theta+\pi}(x)=
\left\{\begin{array}{ll}
f^>_{\theta+\pi}(x)&\forall x>0\;;\\
f^<_{\theta+\pi}(x)&\forall x<0\;;
\end{array}\right.\quad
&\hbox{with}&\quad
f^>_{\theta+\pi}=
{\displaystyle f^>{-}\e^{\ii\theta}\widecheck{f^<}\over\displaystyle2}\ ;
f^<_{\theta+\pi}=
{\displaystyle f^<{-}\e^{-\ii\theta}\widecheck{f^>}\over\displaystyle2}\ .
\end{eqnarray*}

Moreover, $\forall\theta\in[0,2\pi[$, a $\theta$-symmetrical function $f_1$ and
a $\theta+\pi$-symmetrical one $f_2$ are orthogonal:
\[
\langle f_1|f_2\rangle=\langle f_1^<|f_2^<\rangle+\langle f_1^>|f_2^>\rangle
=\langle\widecheck{f_1^>}|\overline{\e^{\ii\theta}}\e^{\ii(\theta+\pi)}|
\widecheck{f_2^>}\rangle+\langle f_1^>|f_2^>\rangle
=-\langle\widecheck{f_1^>}|\widecheck{f_2^>}\rangle+\langle f_1^>|f_2^>\rangle
=0\;.
\]

\section{Complex spectrum of the Coulomb potential}

\begin{figure}[H]
\centering
\includegraphics[width=6cm]{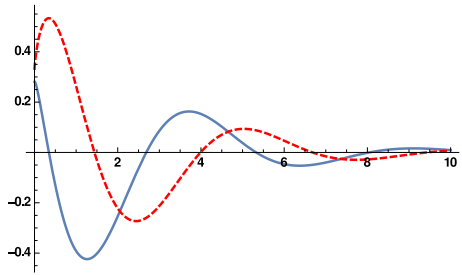}
\caption{Curves of the real (plain line) and imaginary (dashed line)
parts of a function $\phi$, solution of $-\phi''(x)-\phi(x)/x=(1+\ii)\phi(x)$.}
\label{complexEv}
\end{figure}

Contrary to $H_\omega(\Rp)$, $H$ admits states of complex
eigenvalue, included in $L^2(\Rp)$, as for example the following function in
Fig.~\ref{complexEv}. This results from $m_\pm\ne0$\cite{Gitman}.

These solutions are however orthogonal to the quest of self-adjoint extensions
and need not to by studied furthermore.

\section{Self-adjoint extensions of $\R^{3\ast}$}.

If one defines eigenstates in $L^2(\R^{3\ast})$ (the pointed space, where the
position of the fixed charged particle is excluded), one finds that the
classification of $H(\R^{3\ast})$ extensions is equal to that of $H(\Rp)$
ones\cite{Oliveira}.

This exclusion results from the Coulomb potential emitted by the particle, but
the other charged particle is then treated separately from the first one. If one
makes the standard transformation of the two-particle problem into a virtual
charge submitted to an effective Coulomb potential centered at the
barycentre, the exclusion of the barycentre is not founded anymore.

Eventually, this choice would have no influence on the final discussion of
physical states in three dimension, since, as explained in the article, only
Rydberg states have ever been observed.

\section*{Acknowledgments} The author thanks K. Pankrashkin for his
mathematical explanations and clarifications, and A. Jagannathan and J.N.~Fuchs
for their attentive reading.

\begin{table}[H]
\caption{Notations and terminology}
\begin{center}
\begin{tabular}{|l|l|}
\hline
$\Re$ / $\Im$& real/imaginary part of a complex number\\
\hline
$\ii$& the imaginary number. Its conjugate reads
$^{\phantom{\overline{0}}}\overline{\ii}=-\ii$\\
\hline
$\D$ & generic physical space\\
\hline
$\R$ & set of real numbers\\
\hline
$\Rp$ / $\N$ & set of positive real/integer numbers\\
\hline
$E^\ast$& the set $E$ excluding 0 (for any set $E$)\\
\hline
$H$ & hamiltonian\\
\hline
\textbf{simple}& without degeneracy\\
\hline
$\eta$ & adimensional Coulomb parameter\\
\hline
Rydberg states & eigenstates corresponding to $-\eta\in\N$\\
\hline
non Rydberg states & eigenstates corresponding to $-\eta\not\in\N$\\
\hline
$L^1(\D)$ & set of Lebesgue integrable functions defined in $\D$\\
\hline
$L^2(\D)$ & set of Lebesgue square integrable functions defined in $\D$\\
\hline
$L_n$ & Laguerre polynomial\\
\hline
$\cal D$ & generic domain where eigenstates are defined \textbf{for a given
self-adjoint extension}\\
& not to be confused with boundary conditions in real space, applied to
$H(\D)$\\
\hline
$\cal S$ & generic set of \textbf{negative} eigenvalues \textbf{for a given
self-adjoint extension}\\
& (we call it spectrum instead of discrete spectrum)\\
\hline
$\cal B$ & generic set of \textbf{bound} eigenstates \textbf{for a given
self-adjoint extension}\\
\hline
$\cal F$ & generic set of \textbf{free} eigenstates \textbf{for a given
self-adjoint extension}\\
\hline
$\omega$ & parameter which classifies the self-adjoint extensions in the
semi-infinite real line case\\
\hline
$\varpi=(\omega,\theta)$ & parameter which classifies the self-adjoint
extensions in the real line case\\
\hline
\end{tabular}
\end{center}
\label{tabcst}
\end{table}

\end{document}